\documentclass[12pt,onecolumn]{IEEEtran}
\pdfoutput=1
\usepackage{cite}     
\usepackage{graphicx} 
\usepackage{psfrag}   
\usepackage{subfigure}
\usepackage{url}      
\usepackage{stfloats} 
\usepackage{amsmath}  
\usepackage{caption}
\usepackage{array}
\makeatletter
\let\NAT@parse\undefined
\makeatother
\usepackage{times}
\usepackage{epsfig,graphicx,color,pstricks}
\usepackage{amsmath,amssymb,amsthm,amsbsy,amsfonts,latexsym}
\usepackage{mathrsfs}
\usepackage{amsfonts}
\usepackage{amssymb}
\usepackage{enumerate}
\usepackage{bm}
\usepackage{psfrag}
\usepackage{cite}
\usepackage[all]{xy}
\usepackage{subfigure}
\usepackage{graphics}
\usepackage{color}
\theoremstyle{definition}
\newtheorem{def1}{Definition}
\newtheorem{def2}[def1]{Definition}

\theoremstyle{plain}
\newtheorem{thm1}{Theorem}
\newtheorem{thm2}[thm1]{Theorem}
\newtheorem{thm3}[thm1]{Theorem}
\newtheorem{lem1}{Lemma}
\newtheorem{lem2}[lem1]{Lemma}
\newtheorem{thm4}[thm1]{Theorem}
\newtheorem{thm5}[thm1]{Theorem}
\theoremstyle{remark}
\newtheorem{cor1}{Corollary}[thm1]
\newtheorem{cor2}[cor1]{Corollary}
\newtheorem{cor3}[cor1]{Corollary}
\newtheorem{rem1}{Remark}[thm1]
\newtheorem{rem2}{Remark}[thm1]
\newtheorem{rem3}{Remark}[thm1]
\newtheorem{rem4}{Remark}[thm1]

\def\IEEEeqnarraydefcolsep#1#2{\expandafter\def\csname @IEEEeqnarraycolSEP\romannumeral #1\endcsname{#2}%
\expandafter\def\csname @IEEEeqnarraycolSEPDEF\romannumeral #1\endcsname{0}}
\begin{document}
\title{State-Dependent Z Channel}
\author{\IEEEauthorblockN{Saeed Hajizadeh and Mostafa Monemizadeh}}
\maketitle

\begin{abstract}
In this paper we study the ``Z'' channel with side information non-causally available at the encoders. We use Marton encoding along with Gelf'and-Pinsker random binning scheme and Chong-Motani-Garg-El Gamal (CMGE) jointly decoding to find an achievable rate region. We will see that our achievable rate region gives the achievable rate of the multiple access channel with side information and also degraded broadcast channel with side information. We will also derive an inner bound and an outer bound on the capacity region of the state-dependent degraded discrete memoryless Z channel and also will observe that our outer bound meets the inner bound for the rates corresponding to the second transmitter. Also, by assuming the high signal to noise ratio and strong interference regime, and using the lattice strategies, we derive an achievable rate region for the Gaussian degraded Z channel with additive interference non-causally available at both of the encoders. Our method is based on lattice transmission scheme, jointly decoding at the first decoder and successive decoding at the second decoder. Using such coding scheme we remove the effect of the interference completely.
\end{abstract}

\section{Introduction} \label{Intro}
\begin{figure}
\centering
\includegraphics[width=0.4\textwidth]{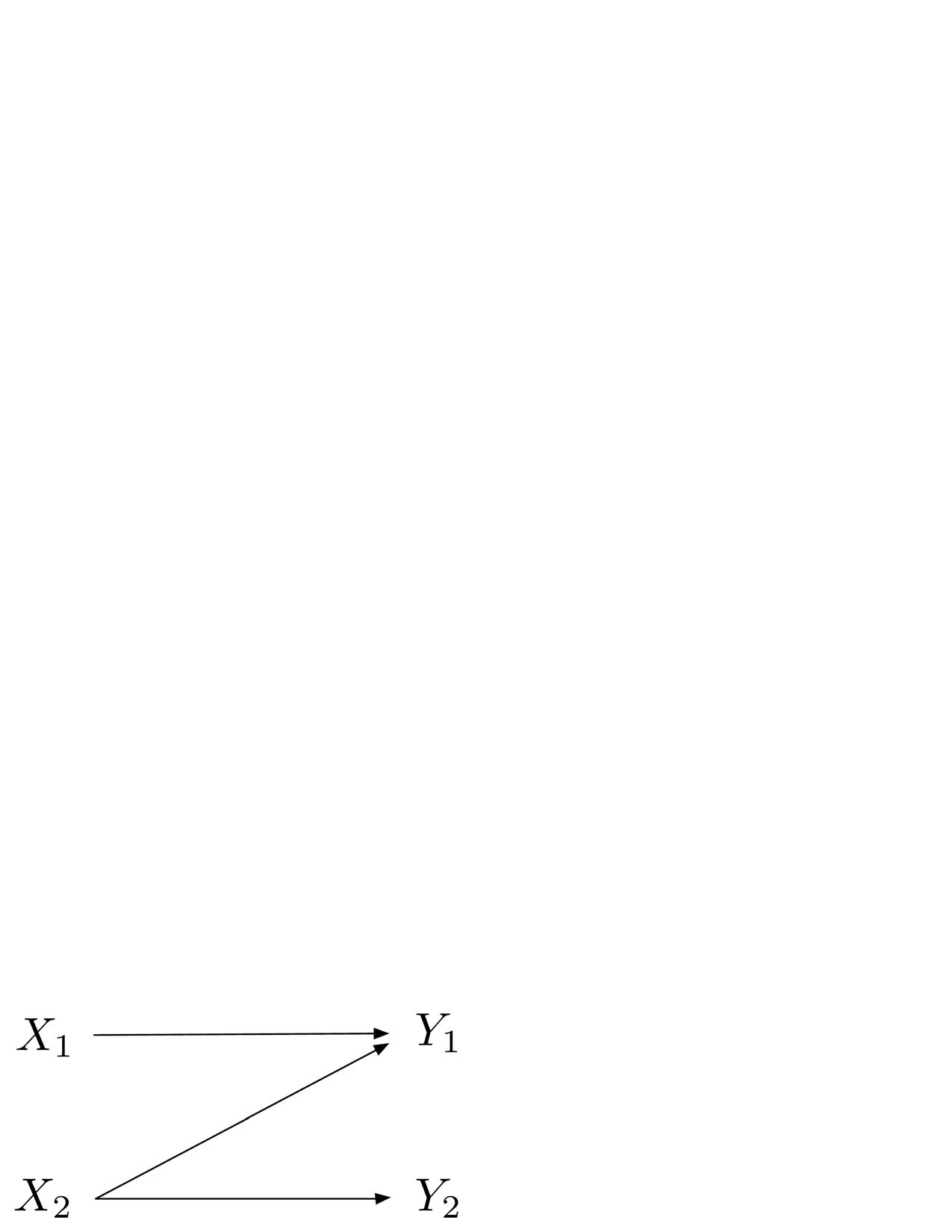}
\centering
\captionsetup{justification=centering}
\caption{The ``Z'' Channel}
\label{Figure1}
\end{figure}
The $Z$ channel is a two-transmitter two-receiver model shown in Fig. \ref{Figure1} where the first sender only wishes to send information to the first receiver whereas the second transmitter sends information to both of the receivers. The $Z$ channel was first studied by Viswanath \emph{et al} \cite{GoldsmithZ:2003} where they introduced the model and found the capacity region of a specialclass of $Z$ channels and the achievable rate of a special case of the Gaussian $Z$ channel (GZC). In \cite{UlukusZ:2004}, Liu and Ulukus obtained several capacity bounds for a class of GZC. Chong-Motani-Garg (CMGE) \cite{CMGZ:2007} studied three different types of degraded Z channel and characterized the capacity region in one type. They also characterized the capacity region of GZC with moderately strong crossover link. 

The capacity region of the general Z channel is still an open problem. The best achievable rate region for the discrete memoryless Z channel until today is due to Do \emph{et al} \cite{SkoglundZ:2011}.

Channels with side information were first studied by Shannon \cite{Shannon:1958} where he characterized the capacity of a point-to-point channel with side information causally available at the transmitters. Gelf'and and Pinsker \cite{gelfand} found the capacity of a single-user channel with side information non-causally available at the encoders. State-dependent multiuser settings have been studied in \cite{three-receiverBC}, \cite{compoundmacISI}, \cite{StateMACFeedbackArXiv}, \cite{ICstate}, and \cite{ExponentialDPC}.

\begin{figure}
\centering
\includegraphics[width=0.9\textwidth]{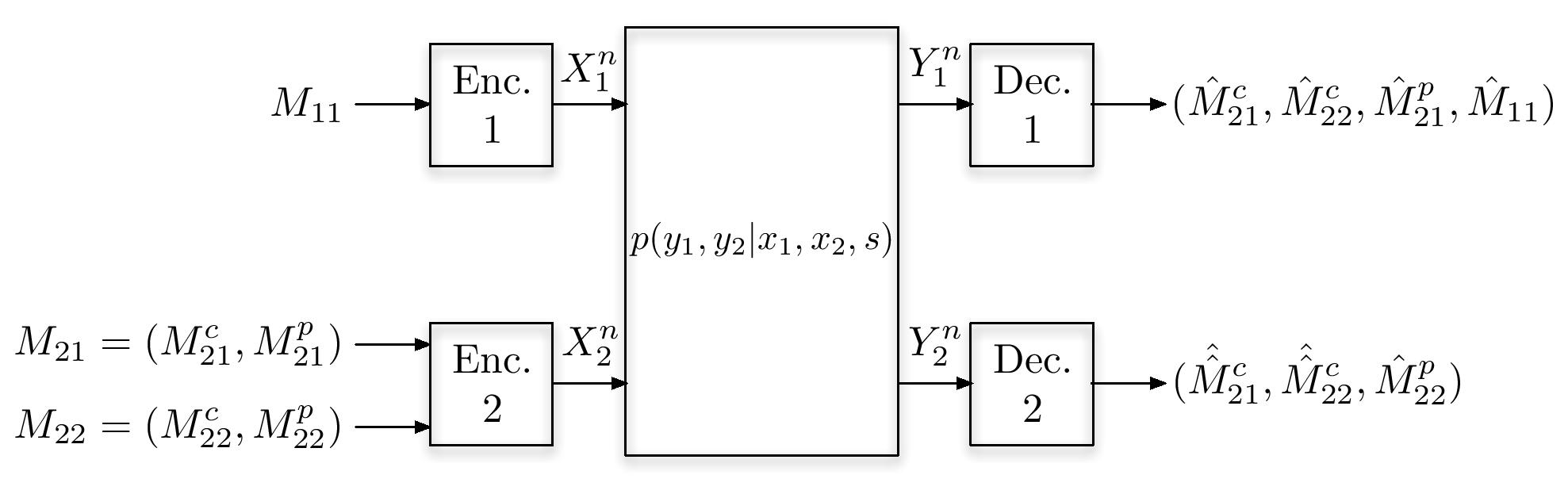}
\centering
\captionsetup{justification=centering}
\caption{The general ``Z'' channel with side information  non-causally available at the encoders.}
\label{Figure2}
\end{figure}

In this paper we study the Z channel with channel state information non-causally available at the encoders that is depicted in Fig. \ref{Figure2}. The reason to study this channel model is buttressed by the applications it has in some wireless communication scenarios such as the case where two communication-involved cells are interfering with each other and thus suffer from a common interference modeled by some S non-causally available to two distinct destination base stations as shown in Fig. 

\begin{figure}
\centering
\includegraphics[width=0.7\textwidth]{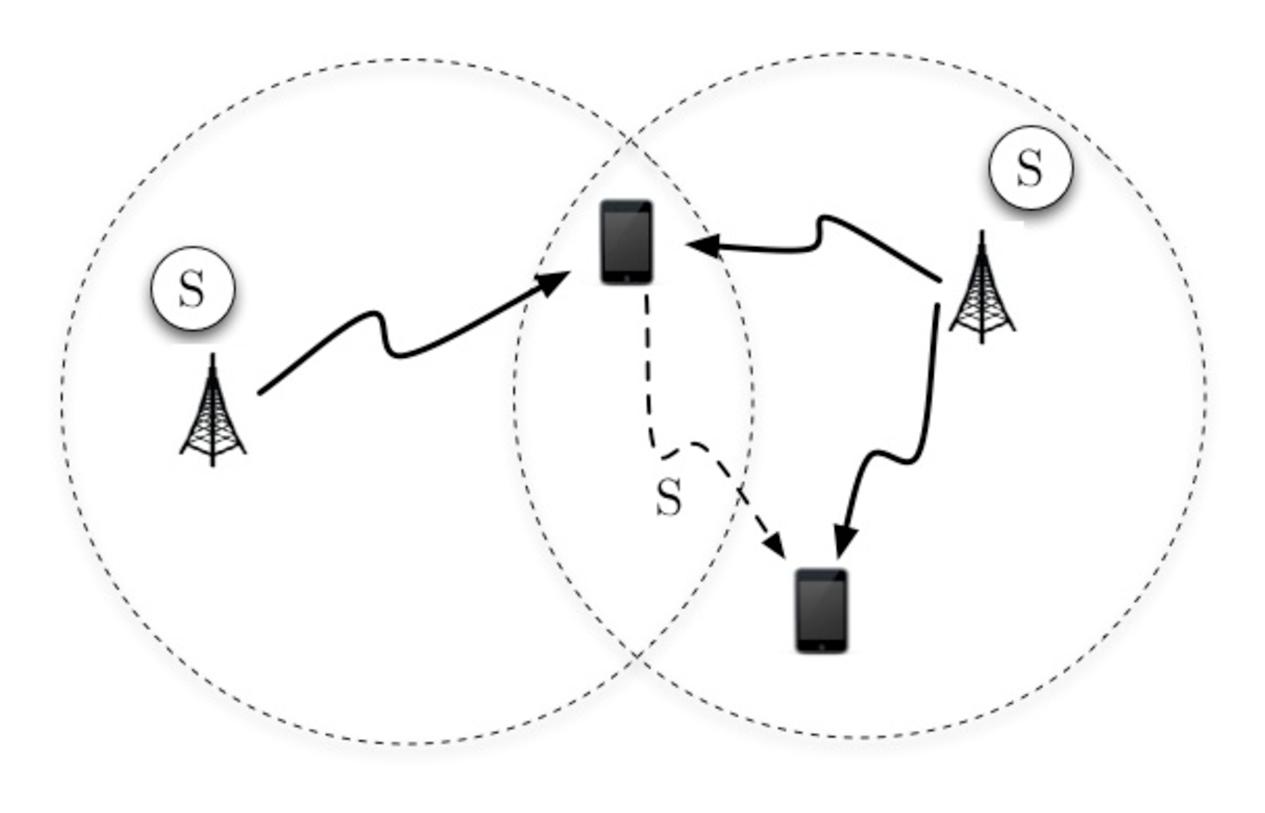}
\centering
\captionsetup{justification=centering}
\caption{A downlink communication scenario as a practical application of the state-dependent Z channel}
\label{Figure3}
\end{figure}
As in Fig. \ref{Figure2}, the first transmitter sends $m_{11} \in \mathscr{M}_{11}=[1:2^{nR_{11}}]$ to $Y_1$ while the second transmitter first splits its messages, $m_{21} \in \mathscr{M}_{21} = [1:2^{nR_{21}}]$ and $m_{22} \in \mathscr{M}_{22} = [1:2^{nR_{22}}]$, to two independent parts, i.e. $M_{21} = (M_{21}^c,M_{21}^p)$ and $M_{22} = (M_{22}^c,M_{22}^p)$ with rates $R_{21} = R_{21}^c + R_{21}^p$ and $R_{22} = R_{22}^c + R_{22}^p$ respectively, and then encodes its messages to send through the channel. The channel state information is non-causally available at the transmitters. The messages $(M_{21}^c,M_{22}^c)$ can be decoded by both receivers, while $M_{2k}^p$ is decoded by its respective receiver, $k = 1,2$.

We propose an achievable rate region using the lattice based coding for the Gaussian degraded ``Z'' channel with additive interference non-causally available at both of the encoders under high-SNR and strong interference regime. Our method is based on lattice transmission scheme, jointly decoding at the first decoder and successive decoding at the second decoder. Using such coding scheme we remove the effect of the interference completely.

The rest of the paper is as follows. In section \ref{defsec}, definitions are provided. In section \ref{mainsec}, we derive an achievable rate region for the general discrete memoryless Z channel with side information non-causally available at the encoders. In section \ref{degsec}, we derive an inner and an outer bound on the capacity region of degraded discrete memoryless Z channel and we will observe that our outer bound coincides with the inner bound for the communication rates of the second transmitter, i.e. $R_{21}$ and $R_{22}$. We will also show that using dirty paper coding, we can remove the negative effect of the interference in the direction of one transmitter-receiver pair in the derived inner bound. In section \ref{latticesec}, we derive an achievable rate region for the Gaussian degraded ``Z'' channel with additive interference non-causally available at both of the encoders using lattice strategies and show that using lattice strategy we can completely remove the interference.The conclusion is given in section \ref{conclusion}.

\section{Definition} \label{defsec}
The discrete memoryless ``Z'' channel with channel state information non-causally available at the transmitter, depicted in Fig. \ref{Figure2}, consists of five finite sets $\mathscr{S}$, $\mathscr{X}_1$, $\mathscr{X}_2$, $\mathscr{Y}_1$, $\mathscr{Y}_2$ and two marginal probability distributions $p(y_2|x_2,s)$ and $p(y_1|x_1,x_2,s)$. The memorylessness nature of the channel imposes the following additional constraint on the channel transition probability
\begin{IEEEeqnarray}{lll}
p(y_1^n,y_2^n|x_1^n,x_2^n,s^n) = \prod_{i=1}^{n}p(y_{2i}|x_{2i},s_i)p(y_{1i}|x_{1i},x_{2i},s_i)
\end{IEEEeqnarray}
A $(2^{nR_{11}},2^{nR_{21}},2^{nR_{22}},n,\epsilon)$ code for the discrete memoryless Z channel with side information consists of two sets of encoding mappings
\begin{IEEEeqnarray}{lll}
e_1:~~~~~~\{1,2,\ldots,2^{nR_{11}}\} \times \mathscr{S}^n \rightarrow \mathscr{X}_1^n \\
e_2:~~~~~~\{1,2,\ldots,2^{nR_{21}}\} \times \{1,2,\ldots,2^{nR_{22}}\} \times \mathscr{S}^n \rightarrow \mathscr{X}_2^n
\end{IEEEeqnarray}
and two sets of decoding mappings
\begin{IEEEeqnarray}{lll}
d_1:~~~~~~\mathscr{Y}_1^n \rightarrow \{1,2,\ldots,2^{nR_{11}}\} \times \{1,2,\ldots,2^{nR_{21}}\} \\
d_2:~~~~~~\mathscr{Y}_2^n \rightarrow \{1,2,\ldots,2^{nR_{22}}\}
\end{IEEEeqnarray}
and an average probability of error defined as the probability that the decoded message does not equal the transmitted message such that
\begin{IEEEeqnarray*}{lll}
p(d_1(y_1^n) \neq (m_{11},m_{21})~\text{or}~d_2(y_2^n) \neq m_{22}) \leq \epsilon
\end{IEEEeqnarray*}
where the messages are assumed to be uniformly distributed on their respective sets.

A rate triple $(R_{11},R_{21},R_{22})$ is said to be achievable for the discrete memoryless ``Z'' channel with side information if there exists a sequence of $(2^{nR_{11}},2^{nR_{21}},2^{nR_{22}},n,\epsilon)$ codes.

\section{The Main Result} \label{mainsec}
In this section, we derive an achievable rate region for the general Z channel with side information. At the first transmitter we apply the Gelf'and-Pinsker random binning and at the second transmitter we use a combination of superposition coding, Marton encoding \cite{marton}, Gelf'and-Pinsker coding, and CMGE \cite{motani} jointly decoding.
\begin{def1}
Define $\mathscr{P}_{ZCSI}$ as the set of all random variables $(S,W,U_0,U_1,U_2,X_1,X_2,Y_1,Y_2)$ such that
\begin{IEEEeqnarray}{lll}
p(s,w,u_0,u_1,u_2,x_1,x_2,y_1,y_2) = \nonumber \\
 p(s) p(w|s) p(x_1|w,s) p(u_0|s) p(u_1,u_2|u_0,s) p(x_2|u_0,u_1,u_2,s) p(y_1,y_2|x_1,x_2,s) \label{gendist}
\end{IEEEeqnarray}
where $(p(x_1|w,s),p(x_2|u_0,u_1,u_2,s)) \in \{0,1\}^2$
\end{def1}
\begin{thm1} \label{theor1}
An achievable rate region for the discrete memoryless Z channel with side information non-causally available at the transmitters, depicted in Fig. \ref{Figure2}, is the closure of the convex hull of the set $\mathscr{R}_{ZCSI} = \bigcup_{p \in \mathscr{P}_{ZCSI}} \mathscr{R}_{ZCSI}(p)$ where
\begin{IEEEeqnarray}{rCl} 
\mathscr{R}_{ZCSI}(p) = \{(R_{11},R_{21},R_{22}):R_{11} &  \leq  & A \label{thm1bound1} \\ 
R_{21} & \leq & B \label{thm1bound2} \\
R_{22} & \leq & C \label{thm1bound3}  \\
R_{11} + R_{21} & \leq & D \label{thm1bound4} \\
R_{21} + R_{22} & \leq & \min\{E,F\} \label{thm1bound5} \\
R_{11} + R_{21} + R_{22} & \leq & \min\{G,H,I\} \label{thm1bound6} \\  \text{\emph{for some}}~(S,W,U_0,U_1,U_2,X_1,X_2,Y_1,Y_2) & \in & \mathscr{P}_{ZCSI} \IEEEnonumber \\ 
\IEEEnonumber \}
\end{IEEEeqnarray}
where we have
\begin{IEEEeqnarray*}{lCl}
A & = & I(W;Y_1|U_0,U_1) - I(W;S|U_0,U_1) \\
B & = & I(U_0,U_1;Y_1|W) - I(U_0,U_1;S|W) \\
C & = & I(U_0,U_2;Y_2) - I(U_0,U_2;S) \\
D & = & I(W,U_0,U_1;Y_1) - I(W,U_0,U_1;S) \\
E & = & I(U_0,U_1;Y_1|W) + I(U_2;Y_2|U_0) - I(U_0,U_1;S|W) - I(U_2;S,U_1|U_0) \\
F & = & I(U_1;Y_1|W,U_0) + I(U_0,U_2;Y_2) - I(U_0,U_1;S|W) - I(U_2;S,U_1|U_0) \\
G & = & I(W,U_1;Y_1|U_0) + I(U_0,U_2;Y_2) - I(W,U_1;S|U_0) - I(U_0;S|W) - I(U_2;S,U_1|U_0) \\
H & = & I(W,U_0,U_1;Y_1) + I(U_2;Y_2|U_0) - I(W,U_0,U_1;S) - I(U_2;S,U_1|U_0) \\
I & = & I(W,U_0,U_1;Y_1) + I(U_0,U_2;Y_2) - I(W,U_0,U_1;S) - I(U_2;S,U_1|U_0) - I(U_0;S)
\end{IEEEeqnarray*}
\end{thm1}
\begin{cor1}
If we put $S \equiv \emptyset$ in \eqref{thm1bound1}-\eqref{thm1bound6}, then we have the achievable rate for the Z channel provided by \cite{SkoglundZ:2011}.
\end{cor1}
\begin{cor2}
If we let no information to be sent to the second receiver, we obtain the achievable rate for the state-dependent multiple access channel with independent sources, i.e. if we set $R_{20}^c = R_{22}^p = 0$, $R_{11} = R_1$, $R_{21}^p = R_2$, $U_0 = U_2 = \emptyset$, $U_1 = U_2^{'}$, $W = U_1^{'}$, and $Y_1 = Y$ in \eqref{midexpr1}-\eqref{midexpr13}, we obtain the closure of the convex hull of the set of all rate-pairs $(R_1,R_2)$ satisfying
\begin{IEEEeqnarray}{rCl}
R_1 & \leq & I(U_1^{'};Y|U_2^{'}) - I(U_1^{'};S|U_2^{'}) \IEEEyessubnumber \\
R_2 & \leq & I(U_2^{'};Y|U_1^{'}) - I(U_2^{'};S|U_1^{'}) \IEEEyessubnumber \\
R_1 + R_2 & \leq & I(U_1^{'},U_2^{'};Y) - I(U_1^{'},U_2^{'};S) \IEEEyessubnumber
\end{IEEEeqnarray}
\end{cor2}
\begin{cor3}
If we set $R_{11} = R_{22}^p = 0$, $R_{20}^c = R_2$, $R_{21}^p = R_1$, and $W = U_2 = \emptyset$, $U_0 = U_2$ in the 12 expressions derived from \eqref{midexpr1}-\eqref{midexpr13} and assuming that receiver $Y_2$ is a degraded version of receiver $Y_1$, we obtain the achievable rate region of the degraded broadcast channel with side information provided in \cite{Steinberg:2005}, namely,
\begin{IEEEeqnarray}{rCl}
R_1 & \leq & I(U_1;Y_1|U_0) - I(U_1;S|U_0) \IEEEyessubnumber \\
R_2 & \leq & I(U_0;Y_2) - I(U_0;S) \IEEEyessubnumber
\end{IEEEeqnarray}
\end{cor3}
\begin{proof}
See Appendix.
\end{proof}

\section{Degraded ``Z'' Channel with Channel State Information} \label{degsec}
Here we determine an inner bound and an outer bound on the capacity region of degraded discrete memoryless Z channel with channel state information. We will see that the derived outer bound coincides with the inner bound on the rates of the second transmitter to the second receiver, i.e. $R_{21}+R_{22}$ and $R_{22}$. Then we show that using dirty paper coding, we can cancel the effect of interference from the direction of one transmitter-receiver pair inside the inner bound provided.
\subsection{An achievable rate region for the degraded discrete memoryless Z channel with channel state information}
\begin{def2} \label{def2}
A ``Z'' channel is said to be degraded if given every $s \in \mathscr{S}$, the following Markov chain holds,
\begin{IEEEeqnarray}{rCl}
X_2 \rightarrow (X_1,S,Y_2) \rightarrow Y_1
\end{IEEEeqnarray}
i.e. the distribution in \eqref{gendist} is limited to the following,
\begin{IEEEeqnarray}{rCl}
p(y_1,y_2|x_1,x_2,s) = p(y_2|x_2,s).p(y_1|x_1,y_2,s).
\end{IEEEeqnarray}
\end{def2}
\begin{thm2} \label{theor2}
The achievable rate region for the degraded ``Z'' channel is the closure of the set of all rate triples $(R_{11},R_{21},R_{22})$ such that
\begin{IEEEeqnarray}{rCl}
R_{11} & \leq & I(W;Y_1|U_0) - I(W;S|U_0) \label{thm2bound1} \\
R_{21} & \leq & I(U_0;Y_1|W) - I(U_0;S|W) \label{thm2bound2} \\
R_{22} & \leq & I(U_2;Y_2|U_0) - I(U_2;S|U_0) \label{thm2bound3} \\
R_{11} + R_{21} & \leq & I(W,U_0;Y_1) - I(W,U_0;S) \label{thm2bound4} \\
R_{21} + R_{22} & \leq & I(U_0,U_2;Y_2) - I(U_0,U_2;S) \label{thm2bound5}
\end{IEEEeqnarray}
\end{thm2}
\begin{proof}
Setting $R_{21}^p = 0$, $R_{20}^c = R_{21}$, $R_{22}^p = R_{22}$, and $U_1 \equiv \emptyset$ in the $12$ expressions derived from \eqref{midexpr1}-\eqref{midexpr13} in the proof of Theorem \ref{theor1}, we can derive \eqref{thm2bound1}-\eqref{thm2bound5}.
\end{proof}
\begin{rem1}
Notice that if the receivers be also aware of the channel state, then using Definition \ref{def2}, one can show that \eqref{thm2bound5} is redundant provided that $x_{1i}$ is a one-to-one mapping.
\end{rem1}
\subsection{An outer bound on the capacity region of the degraded discrete memoryless Z channel with channel state information}
\begin{thm3} \label{theor3}
The set of all rate triples $(R_{11},R_{21},R_{22})$ satisfying,
\begin{IEEEeqnarray}{rCl}
R_{11} + R_{21} & \leq & I(U_0,W;Y_1) - I(W;S) \label{thm3bound1} \\
R_{21} & \leq & I(U_0;Y_1|W,S) \label{thm3bound2} \\
R_{21} + R_{22} & \leq & I(U_0,U_2;Y_2) - I(U_0,U_2;S) \label{thm3bound3} \\
R_{22} & \leq & I(U_2;Y_2|U_0) - I(U_2;S|U_0) \label{thm3bound4}
\end{IEEEeqnarray}
over all distributions of the form 
\begin{IEEEeqnarray}{rCl}
p(s,w,u_0,u_2,x_1,x_2) = p(s)p(w|s)p(u_0,u_2|s)p(x_1|w,s)p(x_2|u_0,u_2,s) \nonumber
\end{IEEEeqnarray}
form an outer bound on the capacity region of the degraded discrete memoryless Z channel with side information.
\end{thm3}
\begin{proof}
See Appendix.
\end{proof}
\begin{rem2}
Notice that achievable rates of the second transmitter coincide with their counterparts in the outer bound and therefore, the second transmitter can communicate optimally with the receivers.
\end{rem2}
\subsection{Achievable rate for the degraded Gaussian Z channel with interference}
Now we study the Gaussian version of the Z channel with channel state information. First we define the Gaussian Z channel model with interference. Then we evaluate the achievable rate found for the discrete memoryless degraded Z channel with side information to the Gaussian case and use dirty paper coding to remove the negative effect of the interference in the channel associated with the first transmitter- receiver pair. The general model of the Gaussian Z channel is as follows,
\begin{IEEEeqnarray}{rCl}
Y_1 & = & X_1 + h_{11}X_2 + h_{12}S + Z_1 \label{Y1Gaussian}  \\
Y_2 & = & X_2 + h_{21}S + Z_2 \label{Y2Gaussian}
\end{IEEEeqnarray}
where for $k=1,2$,
\begin{IEEEeqnarray}{rCl}
\frac{1}{n} \sum_{i=1}^{n} \mathbb{E}(X_k^2) \leq P_k, \qquad Z_k \sim \mathcal{N}(0,N_k),~\text{and}~S \sim \mathcal{N}(0,Q) \nonumber
\end{IEEEeqnarray}
Now we use dirty paper coding presented in \cite{costa} to evaluate the achievable rate provided in Theorem \ref{theor2} for the Additive White Gaussian Noise (AWGN) channel. We first present a Lemma to prove that when Theorem \ref{theor2} is evaluated for the AWGN channel, only one transmitter can successfully cancel the negative effect of the interference while the other transmitter struggles to cancel the interference in its bite of the achievable rate region. Throughout the paper, we optimize the Costa \cite{costa} coefficients so that the first transmitter achieves its own share the non-interfered achievable rate region.

Let $\widetilde{U}_0$, $\widetilde{U}_2$, and $\widetilde{W}$ be three pair-wise independent Gaussian random variables with zero mean and unit variance. Notice that $\widetilde{U}_0$, $\widetilde{U}_2$, and $\widetilde{W}$ are also assumed to be independent of the noise and interference. We also assume that $0 \leq \zeta \leq 1$ is an arbitrary real number and define $\bar{\zeta} \triangleq 1 - \zeta$. Also define,
\begin{IEEEeqnarray}{rCl}
U & \triangleq & \widetilde{U}_0 + \alpha S \nonumber \\
W & \triangleq & \widetilde{W} + \beta S \nonumber \\
U_2 & \triangleq & \widetilde{U}_2 + \gamma S \label{auxG} \\
X_1 & \triangleq & \sqrt{P_1} \widetilde{W} \nonumber \\
X_2 & \triangleq & \sqrt{\zeta P_2}\widetilde{U}_0 + \sqrt{\bar{\zeta} P_2}\widetilde{U}_2 \nonumber 
\end{IEEEeqnarray}
\begin{lem1} \label{lem1}
Using \eqref{Y1Gaussian}, \eqref{Y2Gaussian} and the definition in \eqref{auxG}, we have,
\begin{IEEEeqnarray}{rCl}
I(U_0,W;Y_1) & = & I(U_0,W;Y_1,S) \label{lem1exp1} \\
I(U_0;Y_1|W) & = & I(U_0;Y_1,S|W) \label{lem1exp2} \\
I(W;Y_1|U_0) & = & I(W;Y_1,S|U_0) \label{lem1exp3}
\end{IEEEeqnarray}
provided that,
\begin{IEEEeqnarray}{rCl}
\alpha & = & \frac{h_{11}h_{12}\sqrt{\zeta P_2}}{N_1 + P_1 + h_{11}^2 P_2} \label{costacoeff1} \\
\beta & = & \frac{h_{12}\sqrt{P_1}}{N_1 + P_1 + h_{11}^2P_2} \label{costacoeff2}
\end{IEEEeqnarray}
\end{lem1}
\begin{proof}
See Appendix.
\end{proof}
Now using the definitions in \eqref{auxG}, the equalities \eqref{lem1exp1}-\eqref{lem1exp3} , and the inequalities \eqref{thm2bound1}-\eqref{thm2bound5}, we derive the following Theorem,
\begin{thm4}
The closure of the convex hull of the set of all rate triples $(R_{11},R_{21},R_{22})$ satisfying,
\begin{IEEEeqnarray}{rCl}
R_{11} + R_{21} & \leq & \frac{1}{2} \log \left(1 + \frac{P_1 + h_{11}^2 \zeta P_2}{h_{11}^2 \bar{\zeta} P_2 + N_1}\right) \label{thm4bound1} \\
R_{21} & \leq & \frac{1}{2} \log \left(1 + \frac{h_{11}^2 \zeta P_2}{h_{11}^2 \bar{\zeta} P_2 + N_1}\right) \label{thm4bound2} \\
R_{11} & \leq & \frac{1}{2} \log \left(1 + \frac{P_1}{h_{11}^2 \bar{\zeta} P_2 + N_1} \right) \label{thm4bound3} \\
R_{21} + R_{22} & \leq & \frac{1}{2} \log \left( \frac{P_2 + h_{21}^2 Q + N_2}{\begin{vmatrix}
P_2 + h_{21}^2 Q + N_2 & \sqrt{\zeta P_2} + \alpha h_{21} Q & \sqrt{\bar{\zeta} P_2} + \gamma h_{21} Q \\
\sqrt{\zeta P_2} + \alpha h_{21} Q & 1 + \alpha^2 Q & \alpha \gamma Q \\
\sqrt{\bar{\zeta} P_2} + \gamma h_{21} Q & \alpha \gamma Q & 1 + \gamma^2 Q
\end{vmatrix}} \right) \label{thm4bound4} \\
R_{22} & \leq & \frac{1}{2} \log \left( \frac{\begin{vmatrix}
P_2 + h_{21}^2 Q + N_2 & \zeta P_2 + \alpha h_{21} Q \\
\zeta P_2 + \alpha h_{21} Q & 1 + \alpha^2 Q
\end{vmatrix}}{\begin{vmatrix}
P_2 + h_{21}^2 Q + N_2 & \sqrt{\zeta P_2} + \alpha h_{21} Q & \sqrt{\bar{\zeta} P_2} + \gamma h_{21} Q \\
\sqrt{\zeta P_2} + \alpha h_{21} Q & 1 + \alpha^2 Q & \alpha \gamma Q \\
\sqrt{\bar{\zeta} P_2} + \gamma h_{21} Q & \alpha \gamma Q & 1 + \gamma^2 Q
\end{vmatrix}} \right) \label{thm4bound5}
\end{IEEEeqnarray}
for any $0 \leq \zeta \leq 1$ is an achievable rate region for the degraded Gaussian Z channel.
\end{thm4}
\begin{rem3}
Note that with the Costa coefficients of \eqref{costacoeff1} and \eqref{costacoeff2}, we can cancel the effect of interference from the first three inequalities. In fact, it can be shown that \eqref{costacoeff1} changes if one desires to cancel the effect of interference in the bounds on the rates of the second transmitter.
\end{rem3}
\begin{rem4}
Notice that inequalities \eqref{thm4bound1}-\eqref{thm4bound5} are like those found in Corollary 1 of \cite{Chong2007} where there is no interference.
\end{rem4}

\section{Lattice Strategies for the Gaussian Degraded ``Z'' Channel with Additive Interference} \label{latticesec}
\begin{figure}
\centering
\includegraphics[width=0.8\textwidth]{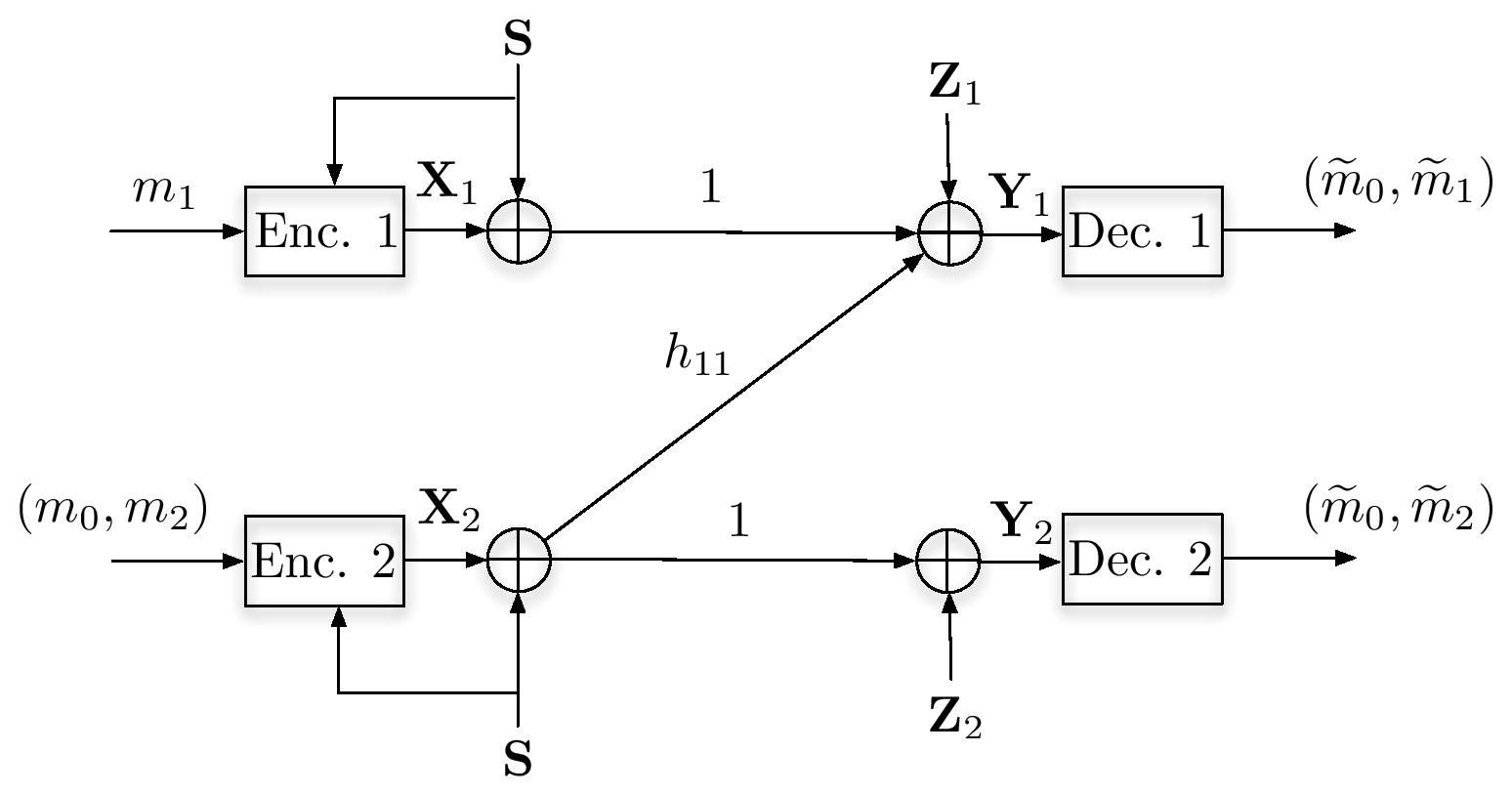}
\centering
\captionsetup{justification=centering}
\caption{The Gaussian degraded ``Z'' channel with additive interference non-causally available at both encoders}
\label{Figure4}
\end{figure}

Now we propose an achievable rate region using lattice based coding for the Gaussian degraded ``Z'' channel with additive interference non-causally available at both of the encoders under the high-SNR and strong interference regime utilizing the standard notation of \cite{Philosof:2011}, \cite{Zamir:1996} and \cite{Forney:2003}. Our method is based on lattice transmission scheme, jointly decoding at the first decoder and successive decoding at the second decoder. Exploiting such a coding scheme we cancel the effect of interference completely. The model of the Gaussian degraded Z channel with the additive interference that we use in this section is depicted in Fig. \ref{Figure4}. The outputs are as follows,
\begin{IEEEeqnarray}{rCl}
Y_1 & = & X_1 + h_{11}X_2 + (1 + h_{11})S + Z_1 \label{lattice-model1} \\
Y_2 & = & X_2 + S + Z_2 \label{lattice-model2}
\end{IEEEeqnarray}
where $h_{11}$ is a real number, $X_i$, $i=1,2$, is the channel input transmitted by user $i$ which is subject to power constraint $P_i$, $Z_i$ is an AWGN with zero mean and variance $N_i$, i.e. $Z \sim \mathcal{N}(0,N_i)$, and the interference signal $S$ is assumed to be $i.i.d$ Gaussian with variance $Q$, i.e. $S \sim \mathcal{N}(0,Q)$, independent of everything else and known non-causally at both encoders.
\begin{thm5}
An achievable rate region for the Gaussian ``Z'' channel with degraded message sets and with side information non-causally available at the transmitters, denoted by $\mathfrak{R}$, is given by
\begin{IEEEeqnarray}{rCl}
\mathfrak{R} & = & \bigcup_{\zeta \in [0,1],\lambda_0 \in \mathbb{R}^+} \mathfrak{R}(\zeta,\lambda_0)
\end{IEEEeqnarray}
with,
\begin{IEEEeqnarray}{rCl}
\mathfrak{R}(\zeta,\lambda_0)  = \Bigg\{ (R_0 \geq 0,R_1 \geq 0,R_2 \geq 0): \nonumber \\
R_0 + R_2 & \leq & \frac{1}{2} \log \left( \left( \frac{ \zeta P_2}{\bar{\lambda}_0^2 \zeta P_2 + \lambda_0^2( \bar{\zeta} P_2 + N_1)} \right) \left( 1 + \frac{\bar{\zeta} P_2}{N_2} \right) \right) \\
R_0 + R_1 + R_2 & \leq & \min \left\{ \frac{1}{2} \log \left( 1 + \frac{P_1}{N_1} \right), \frac{1}{2} \log \left( \frac{h_{11}^2 P_2}{\bar{\lambda}_0^2 h_{11}^2 P_2 + \lambda_0^2 N_1} \right) \right\} \Bigg\}
\end{IEEEeqnarray}
\end{thm5}
\begin{proof}
Our method is based on Lattice transmission scheme, jointly decoding at first decoder and successive decoding at second decoder.

\textbf{\emph{Encoding}}: Consider three lattices $\Lambda_i$, $i=1,2,3$, with fundamental Voronoi regions $\mathcal{V}_i$, and second moments $\sigma_{\Lambda_0}^2 = \zeta P_2$, $\sigma_{\Lambda_1}^2 = P_1$, $\sigma_{\Lambda_2}^2 = \bar{\zeta} P_2$, respectively. Encoder 1 desires to transmit message $m_1$ to receiver 1 while encoder 2 desires to transmit message message $m_0$ to both receivers and message $m_2$ to receiver 2. We let $R_0 = R_{21}$, $R_1 = R_{11}$ and $R_2 = R_{22}$. The message $m_i$ is carried by vector $\boldsymbol{\mathrm{V}}_i$ where $\boldsymbol{\mathrm{V}}_i$ is uniformly distributed over $\mathcal{V}_i$, and $\boldsymbol{\mathrm{V}}_0$, $\boldsymbol{\mathrm{V}}_1$, and $\boldsymbol{\mathrm{V}}_2$ are pair-wise independent. Also let $\boldsymbol{\mathrm{D}}_i$ be uniformly distributed over $\mathcal{V}_i$, i.e. $\boldsymbol{\mathrm{V}}_i \sim \mathcal{U} (\mathcal{V}_i)$, $i=1,2,3$, be three dither signals which are Uniformly distributed over $\mathcal{V}_i$, and independent of each other. In our encoding structure, transmitters 1 and 2 send $\boldsymbol{\mathrm{X}}_1 = \boldsymbol{\mathrm{W}}$ and $\boldsymbol{\mathrm{X}}_2 = \boldsymbol{\mathrm{U}}_0 + \boldsymbol{\mathrm{U}}_2$, respectively, where $\boldsymbol{\mathrm{U}}_0$, $\boldsymbol{\mathrm{W}}$, and $\boldsymbol{\mathrm{U}}_2$ are generated as,
\begin{IEEEeqnarray}{rCl}
\boldsymbol{\mathrm{U}}_0 & = & [\boldsymbol{\mathrm{V}}_0 - \lambda_0 \boldsymbol{\mathrm{S}} + \boldsymbol{\mathrm{D}}_0]~\text{mod}~\mathrm{\Lambda}_0 \\
\boldsymbol{\mathrm{W}} & = & [\boldsymbol{\mathrm{V}}_1 - \lambda_1 \bar{\lambda}_0 \boldsymbol{\mathrm{S}} + \boldsymbol{\mathrm{D}}_1]~\text{mod}~\mathrm{\Lambda}_1 \\
\boldsymbol{\mathrm{U}}_2 & = & [\boldsymbol{\mathrm{V}}_2 - \lambda_2 \bar{\lambda}_0 \boldsymbol{\mathrm{S}} + \boldsymbol{\mathrm{D}}_2]~\text{mod}~\mathrm{\Lambda}_2
\end{IEEEeqnarray}
where $\bar{\lambda}_0 \triangleq 1 - \lambda$ and the MMSE criterion is used to determine $\lambda_i$, $i=0,1,2$. Note that using this encoding structure we have,
\begin{IEEEeqnarray}{rCl}
\frac{1}{n} \mathbb{E} \| \boldsymbol{\mathrm{X}}_k^2 \|  = P_k \qquad k = 1,2. \nonumber
\end{IEEEeqnarray}

\textbf{\emph{Decoding}}: To decode $(\boldsymbol{\mathrm{V}}_0,\boldsymbol{\mathrm{V}}_2)$ at decoder 2, we use a successive decoding scheme in which decoder 2 first decodes $\boldsymbol{\mathrm{V}}_0$ and then decodes $\boldsymbol{\mathrm{V}}_2$. Therefore, receiving $\boldsymbol{\mathrm{Y}}_2$ and using lattice $\Lambda_0$, decoder 2 computes,
\begin{IEEEeqnarray}{rCl}
\boldsymbol{\mathrm{Y}}_0^{(2)} & = & [\lambda_0 \boldsymbol{\mathrm{Y}}_2 - \boldsymbol{\mathrm{D}}_0]~\text{mod}~\Lambda_0 \nonumber \\
& = & [\lambda_0 (\boldsymbol{\mathrm{U}}_0 + \boldsymbol{\mathrm{U}}_2 + \boldsymbol{\mathrm{S}} + \boldsymbol{\mathrm{Z}}_2) - \boldsymbol{\mathrm{D}}_0]~\text{mod}~\Lambda_0 \nonumber \\
& = & [\boldsymbol{\mathrm{V}}_0 - \bar{\lambda}_0 \boldsymbol{\mathrm{U}}_0 + \lambda_0(\boldsymbol{\mathrm{U}}_2 + \boldsymbol{\mathrm{Z}}_2)]~\text{mod}~\Lambda_0 \nonumber \\
& = & [\boldsymbol{\mathrm{V}}_0 + \boldsymbol{\mathrm{Z}}_{02,\epsilon}]~\text{mod}~\Lambda_0
\end{IEEEeqnarray}
where $\boldsymbol{\mathrm{Z}}_{02,\epsilon} = - \bar{\lambda}_0 \boldsymbol{\mathrm{U}}_0 + \lambda_0(\boldsymbol{\mathrm{U}}_2 + \boldsymbol{\mathrm{Z}}_2)$. Therefore, we have,
\begin{IEEEeqnarray}{rCl}
R_0 & = & \frac{1}{n} I(\boldsymbol{\mathrm{V}}_0;\boldsymbol{\mathrm{Y}}_0^{(2)}) \nonumber \\
& = & \frac{1}{n} \left\{ h(\boldsymbol{\mathrm{Y}}_0^{(2)}) - h(\boldsymbol{\mathrm{Y}}_0^{(2)} | \boldsymbol{\mathrm{V}}_0) \right\} \nonumber \\
& = & \frac{1}{n} \left\{ h(\boldsymbol{\mathrm{Y}}_0^{(2)}) - h\left([-\bar{\lambda}_0 \boldsymbol{\mathrm{U}}_0 + \lambda_0 (\boldsymbol{\mathrm{U}}_2 + \boldsymbol{\mathrm{Z}}_2)]~\text{mod}~\Lambda_0\right) \right\} \nonumber \\
& \geq & \frac{1}{2} \log \left( \frac{\zeta P_2}{G (\Lambda_0)}\right) - \frac{1}{2} \log \left( 2\pi e \left( \bar{\lambda}_0^2 \zeta P_2 + \lambda_0^2 (\bar{\zeta} P_2 + N_2) \right)  \right) \label{R0}
\end{IEEEeqnarray}
where \eqref{R0} stems from the fact that,
\begin{enumerate}[i]
\item $\boldsymbol{\mathrm{Y}}_0^{(2)}$ is uniformly distributed over $\mathcal{V}_0$,
\item for a fixed second moment, Gaussian distribution maximizes the entropy,
\item modulo operation reduces the second moment.
\end{enumerate}
and where $G(\Lambda_0)$ is the normalized second moment per dimension of the lattice $\Lambda_0$.
Therefore, as long as $\Lambda_0$ is a good lattice for quantization, we have,
\begin{IEEEeqnarray}{rCl}
R_0 & \leq & \frac{1}{2} \log \left( \frac{\zeta P_2}{\bar{\lambda}_0^2 \zeta P_2 + \lambda_0^2 (\bar{\zeta} P_2 + N_2)} \right) \label{R0sub1}
\end{IEEEeqnarray}
Note that the optimal $\lambda_0$ for transmitter 2 from the second receiver standpoint is $\lambda_0^{\text{opt}-2} = \frac{\zeta P_2}{\zeta P_2 + \bar{\zeta} P_2 + N_2} = \frac{\zeta P_2}{P_2 + N_2}$, and by substituting this $\lambda_0^{\text{opt}-2}$ into \eqref{R0sub1}, we obtain,
\begin{IEEEeqnarray}{rCl}
R_0 & \leq & \frac{1}{2} \log \left( 1 + \frac{\zeta P_2}{\bar{\zeta} P_2 + N_2} \right) \label{R0sub2}
\end{IEEEeqnarray}
Also note that this $\lambda_0^{\text{opt}-2}$ is non-optimal from the first receiver standpoint.

Now using lattice $\Lambda_2$, decoder 2 computes,
\begin{IEEEeqnarray}{rCl}
\boldsymbol{\mathrm{Y}}_2^{(2)} & = & [\lambda_2 (\bar{\lambda}_0 \boldsymbol{\mathrm{Y}}_2 + \boldsymbol{\mathrm{Z}}_{02,\epsilon}) - \boldsymbol{\mathrm{D}}_2] ~\text{mod}~ \Lambda_2 \nonumber \\
& = & [\lambda_2 (\boldsymbol{\mathrm{U}}_2 + \boldsymbol{\mathrm{Z}}_2) + \lambda_2 \bar{\lambda}_0 \boldsymbol{\mathrm{S}} - \boldsymbol{\mathrm{D}}_2] ~\text{mod}~ \Lambda_2 \nonumber \\
& = & [\boldsymbol{\mathrm{V}}_2 - \bar{\lambda}_2\boldsymbol{\mathrm{U}}_2 + \lambda_2 \boldsymbol{\mathrm{Z}}_2] ~\text{mod}~ \Lambda_2
\end{IEEEeqnarray}
Therefore, we have,
\begin{IEEEeqnarray}{rCl}
R_2 & = & \frac{1}{n} I(\boldsymbol{\mathrm{V}}_2;\boldsymbol{\mathrm{Y}}_2^{(2)}) \nonumber \\
& = & \frac{1}{n} \left\{ h(\boldsymbol{\mathrm{Y}}_2^{(2)}) - h(\boldsymbol{\mathrm{Y}}_2^{(2)} | \boldsymbol{\mathrm{V}}_2) \right\} \nonumber \\
& = & \frac{1}{n} \left\{ h(\boldsymbol{\mathrm{Y}}_2^{(2)}) - h([- \bar{\lambda}_2 \boldsymbol{\mathrm{U}}_2 + \lambda_2 \boldsymbol{\mathrm{Z}}_2] ~\text{mod}~ \Lambda_2) \right\} \nonumber \\
& \geq & \frac{1}{2} \log \left( \frac{\bar{\zeta} P_2}{G(\Lambda_2)} \right) - \frac{1}{2} \log \left( 2 \pi e (\bar{\lambda}_2^2 \bar{\zeta} P_2 + \lambda_2^2 N_2) \right) 
\end{IEEEeqnarray}
and with a good lattice for quantization, the achievable $R_2$ is,
\begin{IEEEeqnarray}{rCl}
R_2 & \leq & \frac{1}{2} \log \left( \frac{\bar{\zeta} P_2}{\bar{\lambda}_2^2 \bar{\zeta} P_2 + \lambda_2^2 N_2} \right) \label{R2sub1}
\end{IEEEeqnarray}
Note that the optimal $\lambda_2$ is $\lambda_2^{\text{opt}} = \frac{\bar{\zeta} P_2}{\bar{\zeta} P_2 + N_2}$, and by substituting this $\lambda_2^{\text{opt}}$ into \eqref{R2sub1}, we obtain,
\begin{IEEEeqnarray}{rCl}
R_2 & \leq & \frac{1}{2} \log \left( 1 + \frac{\bar{\zeta} P_2}{N_2} \right)
\end{IEEEeqnarray}

To decode $(\boldsymbol{\mathrm{V}}_0,\boldsymbol{\mathrm{V}}_1)$ at decoder 1, we first set $\zeta = 1$ and $\bar{\zeta} = 0$ and then use a similar method as \cite{Philosof:2011} for MAC. Therefore, as long as $\Lambda_0$ is a good lattice for quantization, we have,
\begin{IEEEeqnarray}{rCl} \label{R0sub3}
R_0 & \leq & \frac{1}{2} \log \left( \frac{h_{11}^2 P_2}{\bar{\lambda}_0^2 h_{11}^2 P_2 + \lambda_0^2  N_1} \right)
\end{IEEEeqnarray}
Note that the optimal $\lambda_0$ for sender 2 from the first receiver standpoint is $\lambda_0^{\text{opt-1}} = \frac{h_{11}^2 P_2}{h_{11}^2 P_2 + N_1}$, and by substituting this $\lambda_0^{\text{opt}-1}$ into \eqref{R0sub3}, we obtain,
\begin{IEEEeqnarray}{rCl}
R_0 & \leq & \frac{1}{2} \log \left( 1 + \frac{h_{11}^2 P_2}{N_1} \right).
\end{IEEEeqnarray}
Also note that this $\lambda_0^{\text{opt}-1}$ is non-optimal from the second receiver standpoint. Similarly, for a good lattice for quantization we have,
\begin{IEEEeqnarray}{rCl} \label{R1sub1}
R_1 & \leq & \frac{1}{2} \log \left( \frac{1}{\bar{\lambda}_1^2 P_1 + \lambda_1^2 N_1} \right).
\end{IEEEeqnarray}
Meanwhile, the optimal $\lambda_1$ is $\lambda_1^{\text{opt}} = \frac{P_1}{P_1 + N_1}$, and by substituting this $\lambda_1^{\text{opt}}$ into \eqref{R1sub1}, we obtain,
\begin{IEEEeqnarray}{rCl}
R_1 & \leq & \frac{1}{2} \log \left( 1 + \frac{P_1}{N_1} \right).
\end{IEEEeqnarray}
\end{proof}

\section{Conclusion} \label{conclusion}
In this paper, we derived an achievable rate region for the general ``Z'' channel with side information non-causally available at the transmitters using Marton encoding, Han-Kobayashi rate splitting, Gelfand-Pinsker coding, and CMGE jointly decoding. We also showed that our rate region subsumes the achievable rate region of the multiple access channel with side information and degraded broadcast channel with side information as its special case. We then derived an inner bound and an outer bound on the capacity region of a special case of degraded ``Z'' channels with side information. We then derived an achievable rate region for the Gaussian ``Z'' channel with degraded message sets and with additive interference non-causally available at both encoders using lattice strategies.

\section{Appendix} \label{appendix}
\subsection{Proof of Theorem \ref{theor1}}
\begin{proof}[Proof of Theorem \ref{theor1}]
Fix a distribution of the form, 
\begin{IEEEeqnarray*}{lll}
p(s)p(w|s)p(x_1|w,s)p(u_0|s)p(u_1,u_2|u_0,s)p(x_2|u_0,u_1,u_2,s)
\end{IEEEeqnarray*}
The second transmitter splits its messages as mentioned in section \ref{Intro}. We then generate the codebook as follows,

Randomly and independently generate $2^{n(R_{11}+\widetilde{R}_{11})}$ sequences $w^n(m_{11},\widetilde{m}_{11})$ each one $i.i.d$ according to $\prod_{i=1}^{n} p(w_i)$ and randomly partition them into $2^{nR_{11}}$ bins. 

Randomly and independently generate $2^{n(R_{21}^c + R_{22}^c + \widetilde{R}_{21}^c + \widetilde{R}_{22}^c)}$ sequences $u^n(m_{21}^c,m_{22}^c,\widetilde{m}_{21}^c,\widetilde{m}_{22}^c)$ each one $i.i.d$ according to $\prod_{i=1}^{n} p(u_{0i})$ and randomly partition them into $2^{n(R_{21}^c + R_{22}^c)}$ bins.

For each pair $(m_{21}^c,m_{22}^c)$, independently generate $2^{n(R_{2k}^p + \widetilde{R}_{2k}^p)}$ sequences $u_k^n(m_{21}^c,m_{22}^c,\widetilde{m}_{21}^c,\widetilde{m}_{22}^c,m_{2k}^p, \widetilde{m}_{2k}^p)$, $k=1,2$, each one $i.i.d$ according to $\prod_{i=1}^{n} p(u_{ki}|u_{0i})$ and randomly partition them into $2^{nR_{2k}^p}$ bins.

\textbf{\emph{Encoding}}: Assume that the transmitters desire to send the triple $(m_{11},m_{21},m_{22})$ with $m_{2k} = (m_{2k}^c,m_{2k}^p)$.\\
TX1, i.e. transmitter 1, looks in bin $m_{11}$ to find some $\widetilde{m}_{11}$ such that $(w^n(m_{11},\widetilde{m}_{11}),s^n)$ is jointly typical. Assume that the chosen index is $\widetilde{M}_{11}$. \\
TX2, in the meantime, looks in bin $(m_{21}^c,m_{22}^c)$ to find some pair $(\widetilde{m}_{21}^c,\widetilde{m}_{22}^c)$ such that the pair,
\begin{IEEEeqnarray}{rCl}
(u^n(m_{21}^c,m_{22}^c,\widetilde{m}_{21}^c,\widetilde{m}_{22}^c),s^n) \nonumber
\end{IEEEeqnarray}
is jointly typical. Assume that the chosen pair is  $(\widetilde{M}_{21}^c,\widetilde{M}_{22}^c)$. \\
TX2 then looks in bin $m_{2k}^p$ to find some $\widetilde{m}_{2k}^p$ such that the pair $u_k^n(m_{21}^c,m_{22}^c, \widetilde{m}_{21}^c, \widetilde{m}_{22}^c,m_{2k}^p,\widetilde{m}_{2k}^p),s^n$ is conditionally jointly typical given $u_0^n$. Given that TX2 has found some $\widetilde{m}_{2k}^p$, that satisfy the above condition, it looks in bin $(m_{21}^p,m_{22}^p)$ to find some $(u_1^n,u_2^n)$ such that the tuple $(u_0^n,u_1^n,u_2^n,s^n)$ is jointly typical. \\
TX1 and TX2 then send $x_{1i} = x_{1i}(w_i,s_i)$ and $x_{2i} = x_{2i}(u_{0i},u_{1i},u_{2i},s_i)$ at time $i$.

\textbf{\emph{Decoding}}: Without loss of generality, assume that the triple $(1,1,1)$ was sent through the channel. The first receiver receives $y_1^n$ and looks for the unique pair $(\widehat{m}_{11},\widehat{m}_{21})$ such that
\begin{IEEEeqnarray}{lll}
(w^n(\widehat{m}_{11},\widetilde{M}_{11}),u_0^n(\widehat{m}_{21}^c,\widehat{m}_{22}^c,\widetilde{M}_{21}^c,\widetilde{M}_{22}^c),u_1^n(\widehat{m}_{22}^c,\widetilde{M}_{21}^c,\widetilde{M}_{22}^c,\widehat{m}_{21}^p,\widetilde{M}_{21}^p),y_1^n) \in A_{\epsilon}^n
\end{IEEEeqnarray}
where $A_{\epsilon}^n$ is the set of jointly typical sequences. \\
The second receiver, meanwhile, receives $y_2^n$ and looks for the unique message index $\widehat{m}_{22}$ such that
\begin{IEEEeqnarray}{lll}
(u_0^n(\widehat{m}_{21}^c,\widehat{m}_{22}^c,\widetilde{M}_{21}^c,\widetilde{M}_{22}^c),u_2^n(\widehat{m}_{22}^c,\widetilde{M}_{21}^c,\widetilde{M}_{22}^c,\widehat{m}_{22}^p,\widetilde{M}_{22}^p),y_2^n) \in A_{\epsilon}^n
\end{IEEEeqnarray}
We now define the following error events for the encoding section,
\begin{IEEEeqnarray}{rCl}
E_1^{enc} & \triangleq & \{(w^n(m_{11},\widetilde{m}_{11}),s^n) \notin A_{\epsilon}^{(n)}~\text{for all}~\widetilde{m}_{11} \in [1:2^{n\widetilde{R}_{11}}] \} \\
E_2^{enc} & \triangleq & \{(u_0^n(m_{21}^c,m_{22}^c,\widetilde{m}_{21}^c,\widetilde{m}_{22}^c),s^n) \notin A_{\epsilon}^{(n)} ~\text{for all}~(\widetilde{m}_{21}^c,\widetilde{m}_{22}^c) \in [1:2^{n\widetilde{R}_{21}^c}] \times [1:2^{n\widetilde{R}_{22}^c}] \} \\
E_{3k}^{enc} & \triangleq & \{(u_0^n(m_{21}^c,m_{22}^c,\widetilde{m}_{21}^c,\widetilde{m}_{22}^c),u_k^n(m_{21}^c,m_{22}^c,\widetilde{m}_{21}^c,\widetilde{m}_{22}^c,m_{2k}^p,\widetilde{m}_{2k}^p),s^n) \notin A_{\epsilon}^{(n)}~\text{for all}~ \widetilde{m}_{2k}^p \in [1:2^{n\widetilde{R}_{2k}^p}] \} \IEEEnonumber \\
&& \:\:\:\:, k=1,2, \\
E_4^{enc} & \triangleq & \{(u_0^n(m_{21}^c,m_{22}^c,\widetilde{m}_{21}^c,\widetilde{m}_{22}^c),u_1^n(m_{21}^c,m_{22}^c,\widetilde{m}_{21}^c,\widetilde{m}_{22}^c,m_{21}^p,\widetilde{m}_{21}^p),u_2^n(m_{21}^c,m_{22}^c,\widetilde{m}_{21}^c,\widetilde{m}_{22}^c,m_{22}^p,\widetilde{m}_{22}^p) \IEEEnonumber \\
&& \:\:\:\:,s^n) \notin A_{\epsilon}^{(n)}~\text{for all}~(\widetilde{m}_{21}^p,\widetilde{m}_{22}^p) \in [1:2^{n\widetilde{R}_{21}^p}] \cup [1:2^{n\widetilde{R}_{22}^p}] \}
\end{IEEEeqnarray}
The decoding error events of the first receiver are defined as follows,
\begin{IEEEeqnarray}{rCl}
E_{11}^{dec} & \triangleq & \{(w^n(1,\widetilde{M}_{11}),u_0^n(1,1,\widetilde{M}_{21}^c,\widetilde{M}_{22}^c),u_1^n(1,1,\widetilde{M}_{21}^c,\widetilde{M}_{22}^c,1,\widetilde{M}_{21}^p),y_1^n) \notin A_{\epsilon}^{(n)} \} \\
E_{12}^{dec} & \triangleq & \{(w^n(\widehat{m}_{11},\widetilde{m}_{11}),u_0^n(\widehat{m}_{21}^c,\widehat{m}_{22}^c,\widetilde{m}_{21}^c,\widetilde{m}_{22}^c),u_1^n(\widehat{m}_{21}^c,\widehat{m}_{22}^c,\widetilde{m}_{21}^c,\widetilde{m}_{22}^c,\widehat{m}_{21}^p,\widetilde{m}_{21}^p),y_1^n) \in A_{\epsilon}^{(n)}~\text{for some} \IEEEnonumber \\
&& \:\:\: (\widehat{m}_{11},\widehat{m}_{21}^c,\widehat{m}_{22}^c,\widehat{m}_{21}^p) \ne (1,1,1,1)~\text{and}~(\widetilde{m}_{11},\widetilde{m}_{21}^c,\widetilde{m}_{22}^c,\widetilde{m}_{21}^p) \ne (\widetilde{M}_{11},\widetilde{M}_{21}^c,\widetilde{M}_{22}^c,\widetilde{M}_{21}^p) \} \\
E_{13}^{dec} & \triangleq & \{(w^n(1,\widetilde{M}_{11}),u_0^n(\widehat{m}_{21}^c,\widehat{m}_{22}^c,\widetilde{m}_{21}^c,\widetilde{m}_{22}^c),u_1^n(\widehat{m}_{21}^c,\widehat{m}_{22}^c,\widetilde{m}_{21}^c,\widetilde{m}_{22}^c,\widehat{m}_{21}^p,\widetilde{m}_{21}^p),y_1^n) \in A_{\epsilon}^{(n)}~\text{for some} \IEEEnonumber \\
&& \:\:\: (\widehat{m}_{21}^c,\widehat{m}_{22}^c,\widehat{m}_{21}^p) \ne (1,1,1)~\text{and some}~(\widetilde{m}_{21}^c,\widetilde{m}_{22}^c,\widetilde{m}_{21}^p) \ne (\widetilde{M}_{21}^c,\widetilde{M}_{22}^c,\widetilde{M}_{21}^p) \} \\
E_{14}^{dec} & \triangleq & \{(w^n(\widehat{m}_{11},\widetilde{m}_{11}),u_0^n(1,1,\widetilde{M}_{21}^c,\widetilde{M}_{22}^c),u_1^n(1,1,\widetilde{M}_{21}^c,\widetilde{M}_{22}^c,\widehat{m}_{21}^p,\widetilde{m}_{21}^p),y_1^n) \in A_{\epsilon}^{(n)}~\text{for some} \IEEEnonumber \\
&& \:\:\: (\widehat{m}_{11},\widehat{m}_{21}^p) \ne (1,1)~\text{and some}~(\widetilde{m}_{11},\widetilde{m}_{21}^p) \ne (\widetilde{M}_{11},\widetilde{M}_{21}^p) \} \\
E_{15}^{dec} & \triangleq &
\{(w^n(1,\widetilde{M}_{11}),u_0^n(1,1,\widetilde{M}_{21}^c,\widetilde{M}_{22}^c),u_1^n(1,1,\widetilde{M}_{21}^c,\widetilde{M}_{22}^c,\widehat{m}_{21}^p,\widetilde{m}_{21}^p),y_1^n) \in A_{\epsilon}^{(n)}~\text{for some}~\widehat{m}_{21}^p \ne 1 \IEEEnonumber \\
&& \:\:\:\: \text{and some}~\widetilde{m}_{21}^p \ne \widetilde{M}_{21}^p \} \\
E_{16}^{dec} & \triangleq &
\{(w^n(\widehat{m}_{11},\widetilde{m}_{11}),u_0^n(1,1,\widetilde{M}_{21}^c,\widetilde{M}_{22}^c),u_1^n(1,1,\widetilde{M}_{21}^c,\widetilde{M}_{22}^c,1,\widetilde{M}_{21}^p),y_1^n) \in A_{\epsilon}^{(n)}~\text{for some}~\widehat{m}_{11} \ne 1 \IEEEnonumber \\
&& \:\:\:\: \text{and some}~\widetilde{m}_{11} \ne \widetilde{M}_{11} \} 
\end{IEEEeqnarray}
The decoding error events for the second receiver are as follows,
\begin{IEEEeqnarray}{rCl}
E_{21}^{dec} & \triangleq & \{(u_0^n(1,1,\widetilde{M}_{21}^c,\widetilde{M}_{22}^c),u_2^n(1,1,\widetilde{M}_{21}^c,\widetilde{M}_{22}^c,1,\widetilde{M}_{22}^p),y_2^n) \notin A_{\epsilon}^{(n)} \} \\
E_{22}^{dec} & \triangleq & \{(u_0^n(m_{21}^c,m_{22}^c,\widetilde{m}_{21}^c,\widetilde{m}_{22}^c),u_2^n(m_{21}^c,m_{22}^c,\widetilde{m}_{21}^c,\widetilde{m}_{22}^c,m_{22}^p,\widetilde{m}_{22}^p),y_2^n) \in A_{\epsilon}^{(n)}~\text{for some} \IEEEnonumber \\
&& \:\:\: (m_{21}^c,m_{22}^c,m_{22}^p) \ne (1,1,1)~\text{and some}~(\widetilde{m}_{21}^c,\widetilde{m}_{22}^c,\widetilde{m}_{22}^p) \ne (\widetilde{M}_{21}^c,\widetilde{M}_{22}^c,\widetilde{M}_{22}^p) \} \\
E_{23}^{dec} & \triangleq & \{(u_0^n(1,1,\widetilde{M}_{21}^c,\widetilde{M}_{22}^c),u_2^n(1,1,\widetilde{M}_{21}^c,\widetilde{M}_{22}^c,m_{22}^p,\widetilde{m}_{22}^p),y_2^n) \in A_{\epsilon}^{(n)}~\text{for some}~m_{22}^p \ne 1~\text{and some} \IEEEnonumber \\
&& \:\:\:\:\widetilde{m}_{22}^p \ne \widetilde{M}_{22}^p \}.
\end{IEEEeqnarray}
Now we bound the probability of encoding the error,
\begin{IEEEeqnarray}{rCl}
p(E_1^{enc}) & = & p(\{(w^n(m_{11},\widetilde{m}_{11}),s^n) \notin A_{\epsilon}^n~\text{for all}~\widetilde{m}_{11} \in [1:2^{n\widetilde{R}_{11}}] \}) \nonumber \\
& = & \prod_{\widetilde{m}_{11}=1}^{2^{n\widetilde{R}_{11}}}p((w^n(m_{11},\widetilde{m}_{11}),s^n) \notin A_{\epsilon}^n) \nonumber \\
& = & \prod_{\widetilde{m}_{11}=1}^{2^{n\widetilde{R}_{11}}}1-p((w^n(m_{11},\widetilde{m}_{11}),s^n) \in A_{\epsilon}^n) \nonumber \\
& \leq & (1-2^{-n(I(W;S)+\delta_1(\epsilon))})^{2^{n\widetilde{R}_{11}}} \nonumber \\
& \leq & e^{-2^{-n(I(W;S)+\delta_1(\epsilon)-\widetilde{R}_{11})}} \leq \frac{\epsilon}{13}, \nonumber 
\end{IEEEeqnarray}
provided that,
\begin{IEEEeqnarray}{rCl} \label{enc-exp1}
\widetilde{R}_{11} \geq I(W;S)+\delta_1(\epsilon)
\end{IEEEeqnarray}
Similarly, we can prove that,
\begin{IEEEeqnarray}{rCl}
p(E_2^{enc}) + p(E_{31}^{enc}) + p(E_{32}^{enc}) + p(E_4^{enc}) \leq \frac{4\epsilon}{13}, \nonumber
\end{IEEEeqnarray}
provided that,
\begin{IEEEeqnarray}{rCl} 
\widetilde{R}_{21}^c + \widetilde{R}_{22}^c & \geq & I(U_0;S) + \delta_2(\epsilon) \label{enc-exp2} \\
\widetilde{R}_{21}^p & \geq & I(U_1;S|U_0) + \delta_3(\epsilon) \label{enc-exp3} \\
\widetilde{R}_{22}^p & \geq & I(U_2;S|U_0) + \delta_4(\epsilon) \label{enc-exp4} \\
\widetilde{R}_{21}^p + \widetilde{R}_{22}^p & \geq & I(U_1;U_2|U_0) + I(U_1,U_2;S|U_0) + \delta_5(\epsilon) \label{enc-exp5}
\end{IEEEeqnarray}
The probability of error at the first receiver is bound as follows, \\
Due to weak law of large numbers $p(E_{11}^{dec}) \leq \frac{\epsilon}{13}$. Other bounds are found as follows,
\begin{IEEEeqnarray}{rCl}
p(E_{12}^{dec}) & = & \sum_{(\widehat{m}_{11},\widehat{m}_{21}^c,\widehat{m}_{22}^c,\widehat{m}_{21}^p) \ne (1,1,1,1)} \sum_{(w^n,u_0^n,u_1^n,y_1^n) \in A_{\epsilon}^{(n)}} p(w^n)p(u_0^n)p(u_1^n|u_0^n)p(y_1^n) \nonumber \\
& \leq & 2^{n(R_{11} + \widetilde{R}_{11} + R_{21}^c + \widetilde{R}_{21}^c + R_{22}^c + \widetilde{R}_{22}^c + R_{21}^p + \widetilde{R}_{21}^p )}.2^{n(H(WU_0U_1Y_1)+\epsilon)}.2^{-n(H(W)-\epsilon)}.2^{-n(H(U_0)-2\epsilon)} \nonumber \\
&& .2^{-n(H(U_1|U_0)-2\epsilon)}.2^{-n(H(Y_1)-\epsilon)} \nonumber \\
& \leq & \frac{\epsilon}{13}
\end{IEEEeqnarray}
provided that,
\begin{IEEEeqnarray}{rCl}
R_{11} + \widetilde{R}_{11} + R_{21}^c + \widetilde{R}_{21}^c + R_{22}^c + \widetilde{R}_{22}^c + R_{21}^p + \widetilde{R}_{21}^p \leq I(W,U_0,U_1;Y_1) + I(U_0;W) + I(W;U_1|U_0) - 7\epsilon \nonumber  \\
\label{dec-exp1}
\end{IEEEeqnarray}
Similarly, one can prove that $\sum_{k=3}^{6}p(E_{1k}^{dec}) \leq \frac{4\epsilon}{13}$ provided that,
\begin{IEEEeqnarray}{rCl}
R_{21}^c + \widetilde{R}_{21}^c + R_{22}^c + \widetilde{R}_{22}^c + R_{21}^p + \widetilde{R}_{21}^p & \leq & I(U_0,U_1;Y_1|W) + I(W;U_0) + I(W;U_1|U_0) \label{dec-exp2} \\
R_{11} + \widetilde{R}_{11} + R_{21}^p + \widetilde{R}_{21}^p & \leq & I(W,U_1;Y_1|U_0) + I(W;U_0) + I(W;U_1|U_0) \label{dec-exp3} \\
R_{21}^p + \widetilde{R}_{21}^p & \leq & I(U_1;Y_1|U_0,W) + I(W;U_0) + I(W;U_1|U_0) \label{dec-exp4} \\
R_{11} + \widetilde{R}_{11} & \leq & I(W;Y_1|U_0,U_1) + I(W;U_0) + I(W;U_1|U_0) \label{dec-exp5}
\end{IEEEeqnarray}
For the second receiver, the analysis of error events imply $p(E_{21}^{dec}) + p(E_{22}^{dec}) \leq \frac{2\epsilon}{13}$ provided that,
\begin{IEEEeqnarray}{rCl}
R_{21}^c + \widetilde{R}_{21}^c + R_{22}^c + \widetilde{R}_{22}^c + R_{22}^p + \widetilde{R}_{22}^p & \leq & I(U_0,U_2;Y_2) \label{dec-exp6} \\
R_{22}^p + \widetilde{R}_{22}^p & \leq & I(U_2;Y_2|U_0) \label{dec-exp7}
\end{IEEEeqnarray}
Therefore,
\begin{IEEEeqnarray}{rCl}
P_e^{(n)} & = & p\left((\cup_{k=1}^{4}E_k^{enc} \cup E_{3k}^{enc}) \cup (\cup_{k=1}^{6} E_{1k}^{dec}) \cup (\cup_{k=1}^{2} E_{2k}^{dec})\right) \leq \frac{5\epsilon}{13} + \frac{6\epsilon}{13} + \frac{2\epsilon}{13} = \epsilon \nonumber
\end{IEEEeqnarray}
Now combining \eqref{dec-exp1}-\eqref{dec-exp7} with \eqref{enc-exp1}-\eqref{enc-exp2}, and setting $R_{20}^c = R_{21}^c + R_{22}^c$ we obtain the following expressions,
\begin{IEEEeqnarray}{rCl}
R_{11} & \leq & A                           \label{midexpr1} \\ 
R_{11} + R_{21}^p & \leq & B                \label{midexpr2} \\
R_{11} + R_{20}^c + R_{21}^p & \leq & C     \label{midexpr3} \\
R_{11} + R_{21}^p + R_{22}^p & \leq & D     \label{midexpr4} \\
R_{11} + R_{20}^c + R_{21}^p + R_{22}^p & \leq & \min\{E,F\} \label{midexpr5} \\
R_{11} + 2R_{20}^c + R_{21}^p + R_{22}^p & \leq & G \label{midexpr6} \\
R_{21}^p & \leq & H \label{midexpr7} \\
R_{22}^p & \leq & I \label{midexpr8} \\
R_{21}^p + R_{22}^p & \leq & J \label{midexpr9} \\
R_{20}^c + R_{21}^p & \leq & K \label{midexpr10} \\
R_{20}^c + R_{22}^p & \leq & L \label{midexpr11} \\
R_{20}^c + R_{21}^p + R_{22}^p & \leq & \min\{M,N\} \label{midexpr12} \\
2R_{20}^c + R_{21}^p + R_{22}^p & \leq & O \label{midexpr13}
\end{IEEEeqnarray}
where,
\begin{IEEEeqnarray}{rCl}
A & = & I(W;Y_1|U_0,U_1) + I(W;U_0,U_1) - I(W;S) \nonumber \\
B & = & I(W,U_1;Y_1|U_0) + I(W;U_0,U_1) - I(U_1;S|U_0) - I(W;S) \nonumber \\
C & = & I(W,U_0,U_1;Y_1) + I(W;U_0,U_1) - I(U_0,U_1;S) - I(W;S) \nonumber \\
D & = & I(W,U_1;Y_1|U_0) + I(U_2;Y_2|U_0) + I(W;U_0,U_1) - I(U_1;U_2|U_0) - I(W;S) - I(U_1,U_2;S|U_0) \nonumber \\
E & = & I(W,U_1;Y_1|U_0) + I(U_0,U_2;Y_2) + I(W;U_0,U_1) - I(U_1;U_2|U_0) - I(W;S) - I(U_0,U_1,U_2;S) \nonumber \\
F & = & I(W,U_0,U_1;Y_1) + I(U_2;Y_2|U_0) + I(W;U_0,U_1) - I(U_1;U_2|U_0) - I(W;S) - I(U_0,U_1,U_2;S) \nonumber \\
G & = & I(W,U_0,U_1;Y_1) + I(U_0,U_2;Y_2) + I(W;U_0,U_1) - I(U_1;U_2|U_0) - I(W;S) - I(U_0;S) \nonumber \\
&& - I(U_0,U_1,U_2;S) \nonumber \\
H & = & I(U_1;Y_1|W,U_0) + I(W;U_0,U_1) - I(U_1;S|U_0) \nonumber \\
I & = & I(U_2;Y_2|U_0) - I(U_2;S|U_0) \nonumber \\
J & = & I(U_1;Y_1|W,U_0) + I(U_2;Y_2|U_0) + I(W;U_0,U_1) - I(U_1;U_2|U_0) - I(U_1,U_2;S|U_0) \nonumber \\
K & = & I(U_0,U_1;Y_1|W) + I(W;U_0,U_1) - I(U_0,U_1;S) \nonumber \\
L & = & I(U_0,U_2;Y_2) - I(U_0,U_2;S) \nonumber \\
M & = & I(U_0,U_1;Y_1|W) + I(U_2;Y_2|U_0) + I(W;U_0,U_1) - I(U_1;U_2|U_0) - I(U_0,U_1,U_2;S) \nonumber \\
N & = & I(U_1;Y_1|W,U_0) + I(U_0,U_2;Y_2) + I(W;U_0,U_1) - I(U_1;U_2|U_0) - I(U_0,U_1,U_2;S) \nonumber \\
O & = & I(U_0,U_1;Y_1|W) + I(U_0,U_2;Y_2) + I(W;U_0,U_1) - I(U_1;U_2|U_0) - I(U_0;S) - I(U_0,U_1,U_2;S) \nonumber
\end{IEEEeqnarray}
The expressions \eqref{midexpr1}-\eqref{midexpr13} first undergo a Fourier-Motzkin procedure using Lemma 1 in \cite{Kramer:2007} after which we will have 12 expressions in \eqref{midexpr1}-\eqref{midexpr13} Now applying the Fourier-Motzkin elimination scheme to the 12 expressions derived from the last step using the constraints $R_{21}+R_{22} = R_{20}^c+R_{21}^p+R_{22}^p$, $R_{21}^p \leq R_{21}$, $R_{22}^p \leq R_{22}$, and non-negativity of the rates, we obtain inequalities \eqref{thm1bound1}-\eqref{thm1bound6}.
\end{proof}
\subsection{Proof of Theorem \ref{theor3}}
Before proving Theorem \ref{theor3}, we need to prove the following Lemma which will be needed throughout the course of the proof procedure,
\begin{lem2} \label{lem2}
Let the discrete memoryless broadcast channel $X \rightarrow (Y_1,Y_2)$ be a less noisy broadcast channel with $Y_2$ being the less noisy receiver than $Y_1$. For every $i=1,2,\ldots,n$, consider $(M,S^{i-1})$ to be any random vector underlying the state-dependent broadcast channel $X \rightarrow (Y_1,Y_2)$. From the discrete memoryless-ness of the channel,
\begin{IEEEeqnarray}{rCl}
(M,S^{i-1}) \rightarrow (X_i,S_i) \rightarrow (Y_{1i},Y_{2i}) \nonumber
\end{IEEEeqnarray}
forms a Markov chain. Then,
\begin{IEEEeqnarray}{rCl}
I(Y_{1,i+1}^{(n)};Y_{1,i}|M,S^{i-1}) \leq I(Y_{2,i+1}^{(n)};Y_{1,i}|M,S^{i-1}) \label{lemexpr}
\end{IEEEeqnarray}
\end{lem2}
\begin{proof}
For any $i+1 \leq r \leq n-1$ we have,
\begin{IEEEeqnarray}{rCl}
I(Y_{1,i+1}^{r},Y_{2,r+1}^{(n)};Y_{1,i}|M,S^{i-1}) & = & I(Y_{1,i+1}^{r-1},Y_{2,r+1}^{(n)};Y_{1,i}|M,S^{i-1}) + I(Y_{1,r};Y_{1,i}|M,S^{i-1},Y_{1,i+1}^{r},Y_{2,r+1}^{(n)}) \nonumber \\
& \stackrel{(a)}{\leq} & I(Y_{1,i+1}^{r-1},Y_{2,r+1}^{(n)};Y_{1,i}|M,S^{i-1}) + I(Y_{2,r};Y_{1,i}|M,S^{i-1},Y_{1,i+1}^{r},Y_{2,r+1}^{(n)}) \nonumber \\
& = & I(Y_{1,i+1}^{r-1},Y_{2,r}^{(n)};Y_{1,i}|M,S^{i-1}) \nonumber
\end{IEEEeqnarray}
where $(a)$ follows from the fact that,
\begin{IEEEeqnarray}{rCl}
(M,S^{i-1},Y_{1,i+1}^{r-1},Y_{2,r+1}^{(n)},Y_{1,i}) \rightarrow (X_r,S_r) \rightarrow (Y_{1,r},Y_{2,r}) \nonumber
\end{IEEEeqnarray}
and also from the fact that $Y_2$ is a less noisy version of $Y_1$. Applying the above inequality a number of times yields the Lemma.
\end{proof}
\begin{proof}[Proof of Theorem \ref{theor3}]
Suppose that $(2^{nR_{11}},2^{nR_{21}},2^{nR_{22}})$ is a code for the degraded discrete memoryless Z channel. We define the auxiliary random variables as follows,
\begin{IEEEeqnarray}{rCl}
W_i & \triangleq & (M_{11},S_{i+1}^{(n)}) \label{auxW} \\
U_{0i} & \triangleq & (M_{21},Y_{2,i+1}^{(n)},S^{i-1}) \label{auxU0} \\
U_{2i} & \triangleq & (M_{21},M_{22},Y_{2,i+1}^{(n)},S^{i-1}) \label{auxU2}
\end{IEEEeqnarray}
We start with the bound on $R_{11}+R_{21}$. We have,
\begin{IEEEeqnarray}{rCl}
n(R_{11}+R_{21}) & \stackrel{(a)}{=} & H(M_{11},M_{21}|Y_1^n) + I(M_{11},M_{21};Y_1^n) - I(M_{11},M_{21};S^n) \nonumber \\
& \stackrel{(b)}{\leq} & n\epsilon_{11n} + \sum_{i=1}^{n}I(M_{11},M_{21};Y_{1i}|Y_{1,i+1}^{(n)}) - I(M_{11},M_{21};S_i|S^{i-1}) \nonumber \\
& \stackrel{(c)}{\leq} & n\epsilon_{11n} + \sum_{i=1}^{n}I(M_{11},M_{21},Y_{1,i+1}^{(n)};Y_{1i}) - I(M_{11},M_{21},S^{i-1};S_i) \nonumber \\
& = & n\epsilon_{11n} + \sum_{i=1}^{n}I(M_{11},M_{21},Y_{1,i+1}^{(n)},S^{i-1};Y_{1i}) - I(S^{i-1};Y_{1i}|M_{11},M_{21},Y_{1,i+1}^{(n)}) \nonumber \\
&& \qquad \qquad ~~~ - I(M_{11},M_{21},S^{i-1},Y_{1,i+1}^{(n)};S_i) + I(Y_{1,i+1}^{(n)};S_i|M_{11},M_{21},S^{i-1}) \nonumber \\
& \stackrel{(d)}{=} & n\epsilon_{11n} + \sum_{i=1}^{n} I(M_{11},M_{21},Y_{1,i+1}^{(n)},S^{i-1};Y_{1i}) - I(M_{11},M_{21},S^{i-1},Y_{1,i+1}^{(n)};S_i) \nonumber \\
& \leq & n\epsilon_{11n} + \sum_{i=1}^{n} I(M_{11},M_{21},S^{i-1};Y_{1i}) + I(Y_{1,i+1}^{(n)};Y_{1,i}|M_{11},M_{21}.S^{i-1}) \nonumber \\
&& \qquad \qquad ~~~ - I(M_{11},M_{21},S^{i-1},Y_{1,i+1}^{(n)};S_i) \nonumber \\
& \stackrel{(e)}{\leq} & n\epsilon_{11n} + \sum_{i=1}^{n} I(M_{11},M_{21},S^{i-1};Y_{1i}) + I(Y_{2,i+1}^{(n)};Y_{1,i}|M_{11},M_{21}.S^{i-1}) \nonumber \\
&& \qquad \qquad ~~~ - I(M_{11},S^{i-1};S_i) \nonumber \\
& \stackrel{(f)}{=} & n\epsilon_{11n} + \sum_{i=1}^{n} I(M_{11},M_{21},S^{i-1},Y_{2,i+1}^{(n)};Y_{1i}) - I(M_{11};S_i|S^{i-1}) \nonumber \\
& \stackrel{(g)}{=} & n\epsilon_{11n} + \sum_{i=1}^{n} I(M_{11},M_{21},S^{i-1},Y_{2,i+1}^{(n)};Y_{1i}) - I(M_{11};S_i|S_{i+1}^{(n)}) \nonumber \\
& \stackrel{(h)}{\leq} & n\epsilon_{11n} + \sum_{i=1}^{n} I(M_{11},M_{21},S^{i-1},S_{i+1}^{(n)},Y_{2,i+1}^{(n)};Y_{1i}) - I(M_{11},S_{i+1}^{(n)};S_i) \nonumber \\
& \stackrel{(j)}{=} & n\epsilon_{11n} + \sum_{i=1}^{n} I(U_{0i},W_i;Y_{1i}) - I(W_i;S_i) \nonumber
\end{IEEEeqnarray}
where $(a)$ follows from the independence of the messages from the state of the channel, $(b)$ follows from Fano's inequality and the chain rule for mutual information, $(c)$ follows from non-negativity of mutual information and from the fact that channel state elements are $i.i.d$, $(d)$ follows from Csiszar-K\"{o}rner identity \cite{Csizar-Kornerbook}, $(e)$ follows from Lemma \ref{lem2} and non-negativity of mutual information, $(f)$ follows from the $i.i.d$-ness of channel state, $(g)$ follows from the two ways that $I(M_{11};S^n)$ can be extended, $(h)$ follows from the $i.i.d$-ness of channel state and non-negativity of mutual information, and $(j)$ follows from \eqref{auxW} and \eqref{auxU0}.
Now we prove the bound on $R_{21}$. We have,
\begin{IEEEeqnarray}{rCl}
nR_{21} & = & H(M_{21}|M_{11},S^n,Y_1^n) + I(M_{21};Y_1^n|M_{11},S^n) \nonumber \\
& \leq & n\epsilon_{12n} + \sum_{i=1}^{n} I(M_{21};Y_{1i}|M_{11},S^{i-1},S_i,S_{i+1}^{(n)},Y_{1,i+1}^{(n)}) \nonumber \\
& \leq & n\epsilon_{12n} + \sum_{i=1}^{n} I(M_{21},S^{i-1},Y_{1,i+1}^{(n)};Y_{1i}|M_{11},S_i,S_{i+1}^{(n)}) \nonumber \\
& \stackrel{(a)}{\leq} & n\epsilon_{12n} + \sum_{i=1}^{n} I(M_{21},S^{i-1};Y_{1i}|M_{11},S_i,S_{i+1}^{(n)}) + I(Y_{2,i+1}^{(n)};Y_{1i}|M_{11},M_{21},S^n) \nonumber \\
& = & n\epsilon_{12n} + \sum_{i=1}^{n} I(M_{21},S^{i-1},Y_{2,i+1}^{(n)};Y_{1i}|M_{11},S_i,S_{i+1}^{(n)}) \nonumber \\
& \stackrel{(b)}{=} & n\epsilon_{12n} + \sum_{i=1}^{n} I(U_{0i};Y_{1i}|S_i,W_i) \nonumber
\end{IEEEeqnarray}
where $(a)$ follows from Lemma \ref{lem2} and $(b)$ follows from \eqref{auxW} and \eqref{auxU0}.
Now we prove the bound on $R_{21}+R_{22}$. We have,
\begin{IEEEeqnarray}{rCl}
n(R_{21}+R_{22}) & = & H(M_{21},M_{22}|Y_2^n) + I(M_{21},M_{22};Y_2^n) - I(M_{21},M_{22};S^n) \nonumber \\
& \leq & n\epsilon_{21n} + \sum_{i=1}^{n} I(M_{21},M_{22},Y_{2,i+1}^{(n)};Y_{2i}) - I(M_{21},M_{22},S^{i-1};S_i) \nonumber \\
& = & n\epsilon_{21n} + \sum_{i=1}^{n} I(M_{21},M_{22},Y_{2,i+1}^{(n)},S^{i-1};Y_{2i}) - I(S^{i-1};Y_{2i}|M_{21},M_{22},Y_{2,i+1}^{(n)}) \nonumber \\
&& \qquad \qquad ~~~ - I(M_{21},M_{22},S^{i-1},Y_{2,i+1}^{(n)};S_i) + I(Y_{2,i+1}^{(n)};S_i|M_{21},M_{22},S^{i-1}) \nonumber \\
& \stackrel{(a)}{=} & n\epsilon_{21n} + \sum_{i=1}^{n} I(M_{21},M_{22},Y_{2,i+1}^{(n)},S^{i-1};Y_{2i}) - I(M_{21},M_{22},S^{i-1},Y_{2,i+1}^{(n)};S_i) \nonumber \\
& \stackrel{(b)}{=} & n\epsilon_{21n} + \sum_{i=1}^{n} I(U_{0i},U_{2i};Y_{2i}) - I(U_{0i},U_{2i};S_i) \nonumber
\end{IEEEeqnarray}
where $(a)$ follows from Csiszar-K\"{o}rner identity \cite{Csizar-Kornerbook} and $(b)$ follows from \eqref{auxU0} and \eqref{auxU2}.\\
We now prove the bound on $R_{22}$. We have,
\begin{IEEEeqnarray}{rCl}
nR_{22} & = & H(M_{22}|M_{21},Y_2^n) + I(M_{22};Y_2^n|M_{21}) - I(M_{22};S^n|M_{21}) \nonumber \\
& \leq & n\epsilon_{22n} + \sum_{i=1}^{n} I(M_{22};Y_{2i}|M_{21},Y_{2,i+1}^{(n)}) - I(M_{22};S_i|M_{21},S^{i-1}) \nonumber \\
& = & n\epsilon_{22n} + \sum_{i=1}^{n} I(M_{22},S^{i-1};Y_{2i}|M_{21},Y_{2,i+1}^{(n)}) - I(S^{i-1};Y_{2i}|M_{21},M_{22},Y_{2,i+1}^{(n)}) \nonumber \\
&& \qquad \qquad ~~~ - I(M_{22},Y_{2,i+1}^{(n)};S_i|M_{21},S^{i-1}) + I(Y_{2,i+1}^{(n)};S_i|M_{21},M_{22},S^{i-1}) \nonumber \\
& \stackrel{(a)}{=} & n\epsilon_{22n} + \sum_{i=1}^{n} I(M_{22},S^{i-1};Y_{2i}|M_{21},Y_{2,i+1}^{(n)}) - I(M_{22},Y_{2,i+1}^{(n)};S_i|M_{21},S^{i-1}) \nonumber \\
& = & n\epsilon_{22n} + \sum_{i=1}^{n} I(M_{21},M_{22},S^{i-1},Y_{2,i+1}^{(n)};Y_{2i}|M_{21},Y_{2,i+1}^{(n)}) \nonumber \\
&& \qquad \qquad ~~~ - I(M_{21},M_{22},Y_{2,i+1}^{(n)},S^{i-1};S_i|M_{21},S^{i-1}) \nonumber \\
& = & n\epsilon_{22n} + \sum_{i=1}^{n} I(S^{i-1};Y_{2i}|M_{21},Y_{2,i+1}^{(n)}) + I(M_{21},M_{22},Y_{2,i+1}^{(n)};Y_{2i}|M_{21},Y_{2,i+1}^{(n)},S^{i-1}) \nonumber \\
&& \qquad \qquad ~~~ - I(Y_{2,i+1}^{(n)};S_i|M_{21},S^{i-1}) - I(M_{21},M_{22},S^{i-1};S_i|M_{21},S^{i-1},Y_{2,i+1}^{(n)}) \nonumber \\
& \stackrel{(b)}{=} & n\epsilon_{22n} + \sum_{i=1}^{n} I(M_{21},M_{22},Y_{2,i+1}^{(n)};Y_{2i}|M_{21},Y_{2,i+1}^{(n)},S^{i-1}) \nonumber \\
&& \qquad \qquad ~~~ - I(M_{21},M_{22},S^{i-1};S_i|M_{21},S^{i-1},Y_{2,i+1}^{(n)}) \nonumber \\
& = & n\epsilon_{22n} + \sum_{i=1}^{n} I(M_{21},M_{22},Y_{2,i+1}^{(n)},S^{i-1};Y_{2i}|M_{21},Y_{2,i+1}^{(n)},S^{i-1}) \nonumber \\
&& \qquad \qquad ~~~ - I(M_{21},M_{22},S^{i-1},Y_{2,i+1}^{(n)};S_i|M_{21},S^{i-1},Y_{2,i+1}^{(n)}) \nonumber \\
& \stackrel{(c)}{=} & n\epsilon_{22n} + \sum_{i=1}^{n} I(U_{2i};Y_{2i}|U_{0i}) - I(U_{2i};S_i|U_{0i}) \nonumber
\end{IEEEeqnarray}
where $(a)$ and $(b)$ both follow from Csiszar-K\"{o}rner identity \cite{Csizar-Kornerbook} and $(c)$ follows from \eqref{auxU0} and \eqref{auxU2}.
\end{proof}
\subsection{Proof of Lemma \ref{lem1}}
First we prove \eqref{lem1exp1}. In fact, it suffices to show that,
\begin{IEEEeqnarray}{rCl}
h(U_0,W|Y_1) & = & h(U_0,W|Y_1,S) \label{lem-toshow}
\end{IEEEeqnarray}
We have,
\begin{IEEEeqnarray}{rCl}
h(U_0,W|Y_1) & = & h(\widetilde{U}_0 + \alpha S, \widetilde{W} + \beta S | X_1 + h_{11}X_2 + h_{12}S + Z_1) \nonumber \\
& = & h(\widetilde{U}_0 + \alpha S, \widetilde{W} + \beta S | \sqrt{P_1} \widetilde{W} + h_{11}\sqrt{\zeta P_2}\widetilde{U}_0 + h_{11}\sqrt{\bar{\zeta} P_2} \widetilde{U}_2 + h_{12}S + Z_1)
\nonumber \\
& = & h(\widetilde{U}-\frac{\alpha}{h_{12}}(\mathbb{T}+Z_1),\widetilde{W}-\frac{\beta}{h_{12}}(\mathbb{T}+Z_1)|\mathbb{T}+h_{12}S+Z_1) \nonumber \\
& = & h(\psi_u,\psi_w|\mathbb{T} + h_{12}S + Z_1) \nonumber
\end{IEEEeqnarray}
where we define,
\begin{IEEEeqnarray}{rCl}
\mathbb{T} & \triangleq & \sqrt{P_1} \widetilde{W} + h_{11}\sqrt{\zeta P_2}\widetilde{U}_0 + h_{11}\sqrt{\bar{\zeta} P_2} \widetilde{U}_2 \nonumber \\
\psi_u & \triangleq & \widetilde{U} - \frac{\alpha}{h_{12}}(\mathbb{T} + Z_1) \nonumber \\
\psi_w & \triangleq & \widetilde{W} - \frac{\beta}{h_{12}}(\mathbb{T} + Z_1) \nonumber
\end{IEEEeqnarray}
One can easily observe that,
\begin{IEEEeqnarray}{rCl}
\mathbb{E} \{ \psi_u \times (\mathbb{T} + Z_1) \} = \mathbb{E} \{ \psi_w \times (\mathbb{T} + Z_1) \} = 0 \nonumber
\end{IEEEeqnarray}
provided that,
\begin{IEEEeqnarray}{rCl}
\alpha & = & \frac{h_{11}h_{12}\sqrt{\zeta P_2}}{N_1 + P_1 + h_{11}^2 P_2} \nonumber \\
\beta & = & \frac{h_{12}\sqrt{P_1}}{N_1 + P_1 + h_{11}^2 P_2} \nonumber
\end{IEEEeqnarray}
and therefore $\psi_u$ and $\psi_w$ are both independent of $\mathbb{T} + Z_1$. We also know that $\psi_u$ and $\psi_w$ are both independent of $S$. Thus we have,
\begin{IEEEeqnarray}{rCl}
h(\psi_u,\psi_w|Y_1) & = & h(\psi_u,\psi_w). \nonumber
\end{IEEEeqnarray}
We can also easily prove that,
\begin{IEEEeqnarray}{rCl}
h(U_0,W|Y_1,S) & = & h(\psi_u,\psi_w). \nonumber
\end{IEEEeqnarray}
and therefore \eqref{lem-toshow} follows. \eqref{lem1exp2} and \eqref{lem1exp3} are proved to hold true in a similar way.
\bibliographystyle{IEEEtran}
\bibliography{refs}

\begin{thebibliography}{10}
\providecommand{\url}[1]{#1}
\csname url@samestyle\endcsname
\providecommand{\newblock}{\relax}
\providecommand{\bibinfo}[2]{#2}
\providecommand{\BIBentrySTDinterwordspacing}{\spaceskip=0pt\relax}
\providecommand{\BIBentryALTinterwordstretchfactor}{4}
\providecommand{\BIBentryALTinterwordspacing}{\spaceskip=\fontdimen2\font plus
\BIBentryALTinterwordstretchfactor\fontdimen3\font minus
  \fontdimen4\font\relax}
\providecommand{\BIBforeignlanguage}[2]{{%
\expandafter\ifx\csname l@#1\endcsname\relax
\typeout{** WARNING: IEEEtran.bst: No hyphenation pattern has been}%
\typeout{** loaded for the language `#1'. Using the pattern for}%
\typeout{** the default language instead.}%
\else
\language=\csname l@#1\endcsname
\fi
#2}}
\providecommand{\BIBdecl}{\relax}
\BIBdecl

\bibitem{GoldsmithZ:2003}
S.~Viswanath, N.~Jindal, and A.~Goldsmith, ``The z channel,'' in \emph{Proc.
  IEEE Global Telecommun. Conf.}, Dec. 2003, pp. 1726--1730.

\bibitem{UlukusZ:2004}
N.~Liu and S.~Ulukus, ``On the capacity region of the gaussian z channel,'' in
  \emph{Proc. IEEE Global Telecommun. Conf.}, Dec. 2004, pp. 415--419.

\bibitem{CMGZ:2007}
H.~F. Chong, M.~Motani, and H.~K. Garg, ``Capacity theorems for the z
  channel,'' \emph{IEEE Trans. Inf. Theory}, vol.~53, pp. 1348--1365, Apr.
  2007.

\bibitem{SkoglundZ:2011}
H.~T. Do, T.~J. Oechtering, and M.~Skoglund, ``Capacity bounds for the z
  channel,'' in \emph{Proc. IEEE Inf. Theory Workshop}, Oct. 2011, pp.
  432--436.

\bibitem{Shannon:1958}
C.~Shannon, ``Channels with side information at the transmitter,'' \emph{{IBM}
  Journal of Res. Development}, vol.~2, pp. 289--293, 1958.

\bibitem{gelfand}
S.~Gel'fand and M.~Pinsker, ``Coding for channels with random parameters,''
  \emph{Probl. Contr. and Inf. Theory}, vol.~9, no.~1, pp. 19--31, 1980.

\bibitem{three-receiverBC}
S.~Hajizadeh and G.~A. Hodtani, ``Three-receiver broadcast channels with side
  information,'' in \emph{International Symposium on Information Theory
  (ISIT)}, 2012, pp. 393 -- 397.

\bibitem{compoundmacISI}
M.~Monemizadeh, S.~A. Seyyedin, G.~A. Hodtani, and S.~Hajizadeh, ``Compound
  multiple access channel with common message and intersymbol interference,''
  in \emph{Telecommunications (IST), Sixth International Symposium on}, 2012,
  pp. 312 -- 316.

\bibitem{StateMACFeedbackArXiv}
H.~N. Andevari, S.~Hajizadeh, B.~Razeghi, and G.~A. Hodtani, ``State-dependent
  multiple access channels with feedback,'' in \emph{23rd Iranian Conference on
  Electrical Engineering (ICEE'15)}, May 2015.

\bibitem{ICstate}
M.~Monemizadeh, G.~A. Hodtani, and S.~Hajizadeh, ``Interference channel with
  common message and slepian-wolf channel state information,'' in
  \emph{Communication and Information Theory (IWCIT), 2013 Iran Workshop on},
  May 2013, pp. 1 -- 6.

\bibitem{ExponentialDPC}
M.~Monemizadeh, S.~Hajizadeh, G.~A. Hodtani, and S.~A. Seyedin, ``Capacity
  bounds for dirty paper with exponential dirt,'' in \emph{Available online at
  http://arxiv.org/abs/1212.3690}.

\bibitem{marton}
K.~Marton, ``A coding theorem for the discrete memoryless broadcast channel,''
  \emph{IEEE Trans. Inf. Theory}, vol.~25, no.~5, pp. 306--311, May 1979.

\bibitem{motani}
H.~Chong, M.~Motani, H.~Garg, and H.~{El Gamal}, ``On the han-kobayashi region
  for the interference channel,'' \emph{IEEE Trans. Inf. Theory}, vol.~54,
  no.~7, pp. 3188--3195, Jul. 2008.

\bibitem{costa}
M.~Costa, ``Writing on dirty paper,'' \emph{IEEE Trans. Inf. Theory}, vol.
  IT-29, pp. 439--441, May 1983.

\bibitem{Chong2007}
H.~Chong, M.~Motani, and H.~Garg, ``{Capacity Theorems for the �Z�
  Channel},'' \emph{IEEE Trans. Inf. Theory}, vol.~53, no.~4, pp. 1348--1365,
  2007.

\end{thebibliography}
\end{document}